%% file: main.tex
\documentclass[sigchi]{acmart}
\usepackage{booktabs} 
\usepackage{enumitem}
\usepackage[normalem]{ulem}

\setcopyright{none}

\begin{document}


\title[A Bayesian Cognition Approach to Improve Data Visualization]{A Bayesian Cognition Approach \\ to Improve Data Visualization}

\author{Yea-Seul Kim}
\affiliation{%
  \institution{University of Washington}
  \city{Seattle}
  \state{Washington}
}
\email{yeaseul1@uw.edu}

\author{Logan A Walls}
\affiliation{%
  \institution{University of Washington}
  \city{Seattle}
  \state{Washington}
}
\email{loganwls@uw.edu}

\author{Peter Krafft}
\affiliation{%
  \institution{University of Washington}
  \city{Seattle}
  \state{Washington}
}
\email{pmkrafft@uw.edu}

\author{Jessica Hullman}
\affiliation{%
  \institution{Northwestern University}
  \city{Evanston}
  \state{Illinois}
  }
\email{jhullman@northwestern.edu}

\renewcommand{\shortauthors}{Kim et al.}

\begin{abstract}
People naturally bring their prior beliefs to bear on how they interpret the new information, yet few formal models exist for accounting for the influence of users' prior beliefs in interactions with data presentations like visualizations. We demonstrate a Bayesian cognitive model for understanding how people interpret visualizations in light of prior beliefs and show how this model provides a guide for improving visualization evaluation. In a first study, we show how applying a Bayesian cognition model to a simple visualization scenario indicates that people's judgments are consistent with a hypothesis that they are doing approximate Bayesian inference. In a second study, we evaluate how sensitive our observations of Bayesian behavior are to different techniques for eliciting people subjective distributions, and to different datasets. We find that people don't behave consistently with Bayesian predictions for large sample size datasets, and this difference cannot be explained by elicitation technique. In a final study, we show how normative Bayesian inference can be used as an evaluation framework for visualizations, including of uncertainty. 

\end{abstract}

%
%




\maketitle

\input{1_intro.tex}
\input{2_relatedWork.tex}

\input{3_study1}

\input{4_study2}

\input{5_study3}
\input{6_discussion}

\bibliographystyle{ACM-Reference-Format}

\end{document}

%% file: 1_intro.tex
\section{Introduction}
Data-driven presentations are used by the media, government, and private sector to inform and influence public opinion. For example, a journalist might present polling data prior to a midterm election in a choropleth map to convey to readers the probability of a Democratic majority in different areas. While visualization designers may acknowledge that users' expectations and prior knowledge (e.g., about the political sentiment within their district, or their own preferences for a given candidate) will influence what they conclude from the visualization, visualization design guidance and evaluation methods rarely acknowledge these factors. Most conventional visualization design guidance implies that finding an effective design means choosing the right combination of visual encodings and comparisons given the target task. Accordingly, evaluations often frame an ideal user as one who minimizes perceptual and other cognitive errors in extracting the information embedded in the visualization. 

Opposing a ``data-only'' view of visualization, models of graphical comprehension from psychologists have described how top-down influences, including prior beliefs and expertise, influence what a person attends to~\cite{carpenter1998}. 
Studies demonstrate how prior knowledge can lead to other ``downstream'' effects on visualization related outcomes, such as how effective an interactive visualization is for different users~\cite{hegarty2004}. Recent visualization experiments demonstrate how eliciting people's beliefs about data directly through the interface~\cite{kim2017,kim2018} can positively impact data recall and prompt critical thinking about data. While this work has examined how people update their beliefs after providing a prediction and seeing others' predictions about data, research in data visualization has yet to develop descriptive or normative cognitive models for predicting and evaluating how people update the prior beliefs they bring upon viewing data. 

Outside of visualization research, psychologists have developed these types of models of how people update their beliefs or opinions about data or a proposition, given information about their prior beliefs~\cite{griffiths2008, griffiths2012}. Bayesian models of cognition compare human cognition, which is assumed to draw on prior knowledge, to a normative standard for rational induction from noisy evidence~\cite{griffiths2006}. By combining key components of Bayesian statistics---including a likelihood function describing the probability of the data given some assumed distribution, a description of the prior probability of different values, and laws of conditional probability--- Bayesian cognitive modeling can \textit{describe} how people update their beliefs given data. Bayesian models have provided explanatory accounts of how people make various real-world perceptual judgments, higher cognitive inferences, and learn and reason inductively~\cite{tenenbaum2006, griffiths2008, lewandowsky2009, griffiths2012}. Bayesian cognitive models can also \textit{prescribe} what updated beliefs are most consistent with one's prior beliefs and the data, providing a normative framework for evaluating interactions with data presentations.

We make several contributions. First, we \textbf{demonstrate a Bayesian cognitive model for assessing how people interpret data presentations} like simple visualizations. In contrast to other frameworks for studying visualization use, a Bayesian cognitive model can be used to examine how people \textit{change their beliefs} in response to presented data.

Deploying the model we develop, we \textbf{characterize the extent to which the belief updating of users of a simple visualization of survey results resembles Bayesian inference} (Study 1). We find evidence that on average people update their beliefs rationally, but individuals often deviate from expectations of rational belief updating. These findings hold across multiple datasets and prior elicitation methods (Study 2). We find that people deviate considerably more from the predictions of Bayesian inference even in aggregate when presented with datasets of a very large sample size. 

Finally, we \textbf{demonstrate how a Bayesian cognitive model can be used to evaluate data presentations} (Study 3). We show how Hypothetical Outcome Plots (HOPs), animated plots that show uncertainty via draws from a distribution, improve deviation from normative Bayesian responses relative to not presenting error information.

%% file: 2_relatedWork.tex
\section{Background}

\subsection{Interpreting Data Presentations} 
Cognitive psychologists proposed early models of visualization interpretation implying that ``top down'' factors relating to a user's information needs, prior knowledge, and graph literacy affect how visualized data is interpreted, for example, by guiding attention~\cite{kosslyn1989,shah1999}. Studies in graph comprehension provide evidence of such top-down effects~\cite{canham2010, pinker1990,zacks1999, padilla2018}. For example, static visualization of processes, which require use of internal representations to interpret, often outperform animations~\cite{hegarty2004, mayer2014}. Other studies show that externalizing one's internal representations leads to better understanding of visualized information~\cite{cox1999,cosmides1996,hegarty1997,natter2005,stern2003}. 

\subsubsection{The Impact of Prior Knowledge \& Beliefs}
In reflecting on the ``value'' of visualization, Van Wijk notes that the knowledge gained from a visualization will depend on the prior knowledge that a user brings~\cite{van2005}. Recent research demonstrates that while visualizations are slightly more likely to persuade people to change their attitudes about a data driven topic (e.g., to be more likely to believe that some factor X causes some symptom Y), the polarity of the person's original attitude influences the strength of the visualization's effect~\cite{pandey2014}.
Going a step further toward understanding the role of prior beliefs, Kim et al. 
show that asking visualization users to ``draw'' their predictions in an interactive visualization prior to seeing the observed data can help them remember data ~30\% better~\cite{kim2017,kim2018}, perhaps by increasing their ability to compare the observed data to their expectations.
Deviation between a predicted trend or value and the observed trend or value has been shown to be predictive of people's updated beliefs and ability to recall data~\cite{kim2018,munnich2007}.
However, these works focused on eliciting a user's single best prediction of a trend, rather than a distribution over possible values, which is required for making use of the normative predictors that are possible from a Bayesian approach.

\subsection{Bayesian Cognition}

In cognitive science, Bayesian statistics has proven to be a powerful tool for modeling human cognition~\cite{griffiths2008, griffiths2012}. In a Bayesian framework, individual cognition is modeled as Bayesian inference: an individual is said to have implicit beliefs about the world ("priors"); when the individual observes new data, their prior is "updated" to produce a new set of beliefs which account for the observed data (this new set of beliefs is referred to as the "posterior"). The prior is formalized as a probability distribution and Bayes' rule is used to obtain the posterior from the prior distribution and the likelihood function that the observed data is derived from. 

This approach has been used to model many aspects of human cognition at various levels of complexity, such as object perception~\cite{kersten2003}, causal reasoning~\cite{steyvers2003}, and knowledge generalization~\cite{tenenbaum2006}. 

Particularly relevant to our work is a study conducted by Griffiths and Tenenbaum~\cite{griffiths2006}, which compared people's predictions for a number of everyday quantities to the predictions made by a model that used the empirical distribution as a prior (e.g., for human lifespans they used a model with a prior calculated from historical human lifespan data). The study found that although there was variance between individuals, in aggregate people's judgments closely resembled the normative Bayesian posterior. We are similarly interested in how judgments that people make in everyday interactions with data presentations (like visualizations) compare to the expectations of normative Bayesian inference.

\subsubsection{Approximate Inference \& Sampling Behavior}
While Bayesian models of cognition have seen wide applications, the idea that human cognition is accurately described as Bayesian inference is inconsistent with previous influential findings in cognitive psychology from authors such as Tversky and Kahneman~\cite{tversky1974}. Tversky and Kahneman found evidence that humans often use simple heuristics in their decisions, and that these heuristics lead to sub-optimal judgments. More recent research indicates that heuristics are adaptive and often lead to accurate judgments (e.g.,~\cite{goldstein2009}). A recently proposed explanation which reconciles the seemingly opposing findings between Bayesian models of cognition and the idea that heuristics lead to non-optimal judgments is motivated by Bayesian cognition \cite{griffiths2015rational}: what if human cognition is not \textit{exact} Bayesian inference, but instead is \textit{approximate} Bayesian inference?
One such approach proposes that while people have a prior probability distribution which encodes their beliefs, they do not form judgments using the entire distribution at once \cite{vul2014}. Instead, they take a small number of samples from the distribution, and reason with these samples instead of the full distribution (we which refer to as \textit{sample-based Bayesian}). Being a sample-based Bayesian can lead to sub-optimal individual inferences, but in aggregate, it can produce results very similar to exact Bayesian inference.

\subsubsection{Application to Data Visualization}
Recent work by Wu et al.~\cite{wu2017} explored the application of the Bayesian framework to examine how people update their beliefs when viewing visualized data. 
However, Wu et al. prompted participants to internalize a provided prior, show them the observed data, and then ask for their posterior beliefs. Using a fixed prior is not ideal in cases where participants' pre-existing beliefs about a phenomenon will impact their ultimate beliefs. Our work demonstrates how to elicit and model participants' personalized priors for a more realistic application of Bayesian inference, including proposing and evaluating multiple elicitation techniques.

\begin{figure}[t]
 \centering
  \includegraphics[width=\columnwidth]{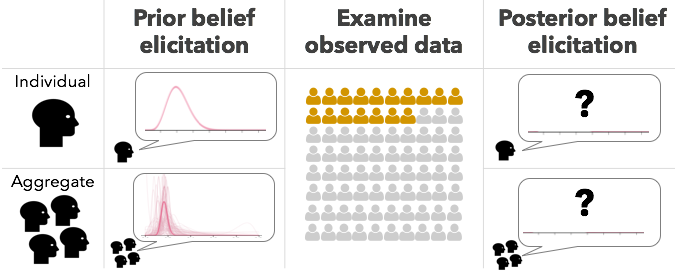}
  \Description{Bayesian inference at individual \& aggregate level. }
 \caption{Bayesian inference at individual \& aggregate level. 
 }
\vspace{-0.1in}
\label{fig:bayesian_update}
\end{figure}

\section{Developing Research Questions \& Goals}

Prior beliefs clearly play a role in data interpretation. However, belief updating is rarely formally modeled in research related to data presentation and visualization. Studies of Bayesian cognition suggest that Bayesian inference can be used to characterize many aspects of learning and cognition. We apply a Bayesian cognitive modeling approach to a simple data interpretation task to understand where people align with, and deviate from, normative Bayesian inference individually and in aggregate. While the computational complexity of Bayesian inference makes it doubtful that cognition uses exact inference~\cite{kahneman2011}, in the context of interpreting presented data in everyday settings (such as in data journalism) we would expect under Bayesian assumptions to see that people (1) are capable of providing \textit{priors} describing the uncertainty in their beliefs about a parameter, and (2) update these beliefs to incorporate observed data. Our work sheds light on the degree to which these assumptions hold for a simple data interpretation task.

In contrast to prior work in Bayeisan cognition that avoids obtaining priors directly from people~\cite{griffiths2006,wu2017}, we design and apply a paradigm in which we \textit{elicit people's prior and posterior beliefs about the probability that a parameter takes various values} (Studies 1, 2, 3). 
Though it is commonly argued that people have difficulties reasoning about probability, the notion that people are capable of maintaining subjective probabilities is well-established in decision theory, congruent with canonical work in judgment and decision making like that of Tversky and Kahneman~\cite{tversky1974}, and supported by a body of work in economics on subjective probability elicitation, including from laypeople (see~\cite{manski2018} for a review. Having obtained prior beliefs, we fit a distribution to them then use Bayes' rule to compute the \textit{normative posterior distribution} for each person, the posterior distribution that is expected if the person is a perfect Bayesian agent given the observed data and their prior distribution. 

In a first study, we compare the distribution fit to the posterior beliefs we elicited from each person to the normative posterior beliefs computed using that person's prior Bayesian solutions (Fig.~\ref{fig:bayesian_update} top row). We also compare people's \textit{aggregate posterior distribution} (i.e. the posterior distributions representing the aggregate of all people's posterior distributions) to the \textit{normative aggregate posterior distribution} (i.e., the normative posterior distribution calculated using a prior distribution representing the aggregate of all people's prior distributions)   (Fig.~\ref{fig:bayesian_update} bottom row). Alignment between people's responses and the normative Bayesian solution at this aggregate level may suggest that people are ``sample-based Bayesians''~\cite{vul2014} performing approximately inference.

Prior work in visualization and judgment and decision making suggests that different subjective probability elicitation techniques can produce varying results, perhaps because some techniques (such as frequency framings) better align with people's internal representations of uncertainty~\cite{goldstein2014,hullman2018,o2006}. In a second study, we assess \textit{how sensitive people's responses are to different elicitation methods}, which vary in the input format for beliefs they use (i.e., continuous probability versus discrete samples). 

In a third study, we show how a Bayesian cognitive model can be used to assess the effectiveness of design changes. One aspect of visualization design that is likely to be relevant to how people update beliefs is the presentation of uncertainty. If people see the observed data as more certain than it is (e.g., reflecting belief in the law of small numbers~\cite{tversky1971}), their posterior judgments may reflect overweighting of the observed data and underweighting of their prior. 
On the other hand, if people see the data as less certain that it is (e.g., non-belief in the law of large numbers~\cite{benjamin2016}), their posterior judgments may reflect underweighting of the observed data and overweighting of their prior.
To demonstrate how a Bayesian cognitive model can support visualization design decisions, we \textit{compare the results of Bayesian modeling across a default static visualization typical of those found in the media and an animated hypothetical outcome plot (HOP~\cite{hullman2015}) uncertainty visualization}.

%% file: 3_study1.tex
\section{S1: Bayesian Data Interpretation}

We demonstrate how a Bayesian model cognition of cognition can be used as a framework for assessing visualization interpretation.
We evaluate the extent to which individuals' judgments are consistent with people being ``fully'' Bayesian by assessing how closely their individual posterior distributions align with the normative posterior distribution calculated given their prior. 
Secondly, we also consider whether people's judgments might instead be consistent with people being ``sample-based'' Bayesians (one form of being approximately Bayesian) by evaluating how closely the aggregate posterior distribution aligns with the normative aggregate posterior distribution. 

\subsection{Study Design}
We designed a between-subjects experiment with 50 participants. We determined sample size via a prospective power analysis conducted on pilot data with a desired power of at least 0.8 ($\alpha$=0.05) to detect a difference between the normative distribution and aggregated posterior distribution.
We recruited participants from Amazon Mechanical Turk (AMT), rewarding their participation with \$1.0. The average time to complete the task was 7.3 minutes (SD=5.2).

\subsubsection{Dataset \& Presentation}
For the purposes of understanding how Bayesian cognitive modeling might provide insight into visualization interpretation, we sought a realistic yet relatively simple dataset similar those shown in the media or public facing reports. We selected a dataset with a single variable which represents a proportion. The dataset describes survey results intended to measure attitudes towards mental health in the tech workplace (N=747)~\cite{mentalsurvey}. We chose one question from the survey ``how often do you feel that mental health affects your work?'' to formulate our proportion parameter: ``the proportion of women in the tech industry who feel that mental health affects their work often.'' To present the observed proportion to participants in our study, we created an ``info-graphic'' style visualization (Fig.~\ref{fig:vis} (a)) which shows this proportion using a grid format commonly used in the media to present proportions (e.g.,~\cite{gridexample3, gridexample2, gridexample1}).  

\subsubsection{Prior \& Posterior Elicitation}
\begin{figure*}[htb]
 \centering
  \includegraphics[width=0.78\textwidth]{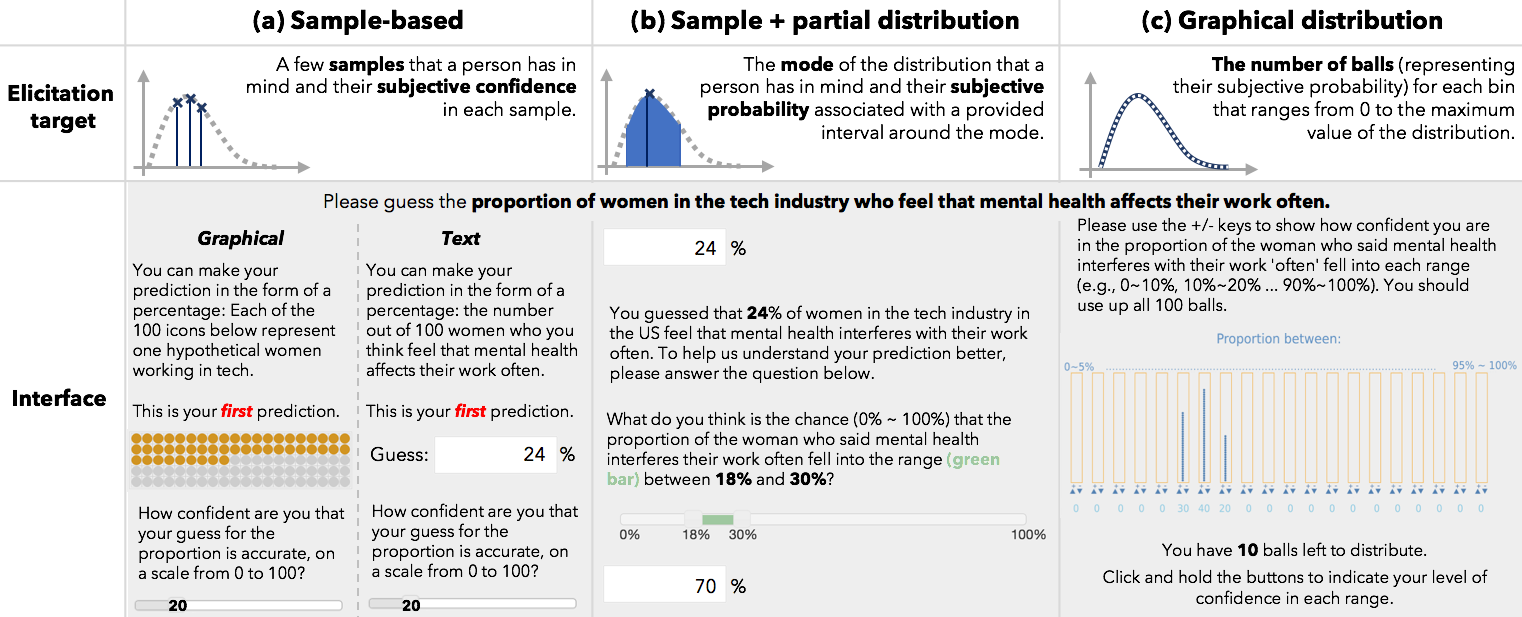}
  \Description{Elicitation target and interface associated with each elicitation technique.}
 \caption{Elicitation target and interface associated with each elicitation technique. We developed two sample-based techniques (a), and used an interval technique~\cite{wu2008} (b) and a graphical ``balls and bins'' technique~\cite{goldstein2014} (c) from the literature.} 
\vspace{-0.1in}
\label{fig:s2_technique}
\end{figure*}

To elicit participants' prior and posterior distributions, we used a technique that asks participants about two properties of their internal distribution: the most probable value of the parameter (\textit{mode (m)}) and their \textit{subjective probability} (Fig.~\ref{fig:s2_technique}(b)) that the parameter falls into the interval around the mode ([$m - 0.25m, m + 0.25m$]). Prior research in probability elicitation for proportions indicates that this technique is less sensitive to noise which arises from externalizing subjective uncertainty compared to other techniques ~\cite{wu2008}. A second benefit of this approach is that estimates of Beta distribution parameters can be analytically computed from participants' answers~\cite{fox1966}.

\subsection{Results}

\subsubsection{Fitting Individual Responses}
We first converted participants' elicited responses of prior and posterior beliefs to Beta distributions using an optimization approach suggested in previous work~\cite{o2006}. The approach finds an optimal Beta distribution parameterized by $\alpha$ and $\beta$ which minimizes the sum of two terms: (1) the square difference between the participants' mode and the estimated mode of the Beta distribution and (2) the square difference between the probability that each participant associated with the interval and the estimated probability of the interval in the distribution being optimized.

\subsubsection{Fitting Aggregate Responses}
To obtain parameters for the aggregated prior/posterior distributions ($\alpha_{\text{agg}}$ and $\beta_{\text{agg}}$), we averaged participants' $\alpha$s and $\beta$s respectively from the individual prior/posterior distributions: $\alpha_{\text{agg}}=(\alpha_{\text{1}}+...+\alpha_{\text{N}}) / N, \beta_{\text{agg}}=(\beta_{\text{1}} + ... + \beta_{\text{N}}) / N$   (where $N = \#$ of participants).

\subsubsection{Calculating Normative Posteriors} We can calculate a participant's normative posterior by using $\alpha$ and $\beta$ estimates from their prior distribution combined with the number of successes (e.g., the number of women who said their mental health affects their work often) and failures (e.g., the number of women who said their mental health affects their work not often) in the observed data (Eq.~\ref{eq:normative}). The $\alpha$ and $\beta$ for the aggregated normative posterior are calculated in the same manner using the aggregated prior $\alpha$ and $\beta$ estimates. 

\vspace{-2mm}
\begin{equation}
  \label{eq:normative}
  \begin{aligned}
    \alpha_{\text{normative posterior}}  = \# successes + \alpha_{\text{prior}}\\        
    \beta_{\text{normative posterior}}  = \# failures + \beta_{\text{prior}} 
  \end{aligned}
\end{equation}


\begin{figure}[htb]
 \centering
  \includegraphics[width=0.35\textwidth]{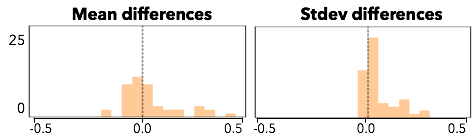}
  \Description{Distributions of residuals (observed - predicted) for participants' posteriors' means and standard deviations and the means and standard deviations of the normative posteriors.}
 \caption{Distributions of residuals (observed - predicted) for participants' posteriors' means and standard deviations and the means and standard deviations of the normative posteriors.}
\vspace{-0.1in}
\label{fig:s1_diff_mean_stdev}
\end{figure}

We evaluate the degree to which individual and aggregate posterior distributions resemble the normative Bayesian posterior distributions by plotting residuals (\textit{observed - predicted}) when predicting the means and standard deviations of participants' posterior distributions using normative Bayesian inference (Fig.~\ref{fig:s1_diff_mean_stdev}).
A distribution of residuals that is loosely centered around zero suggests ``noisy'' Bayesian inference, where each individual may deviate from the normative posterior due to approximate inference but in aggregate, the observed posterior resembles the normative posterior. Residuals for means are roughly centered around zero, with 95\% of the values falling between -0.16 and 0.58). 
A small number of participants provided means that were much greater than predicted (i.e., believed that the true proportion of women in tech who feel that mental health affects their work often was much larger than predicted from the prior and the observed data).   
Residuals for standard deviation are also roughly distributed around zero, but show a stronger bias toward larger standard deviation in one's posterior judgments. 
This suggests a tendency among participants to provide posterior beliefs indicating more uncertainty than is rational given the observed data according to Bayesian inference.

\begin{figure}[htb]
 \centering
  \includegraphics[width=3.5in]{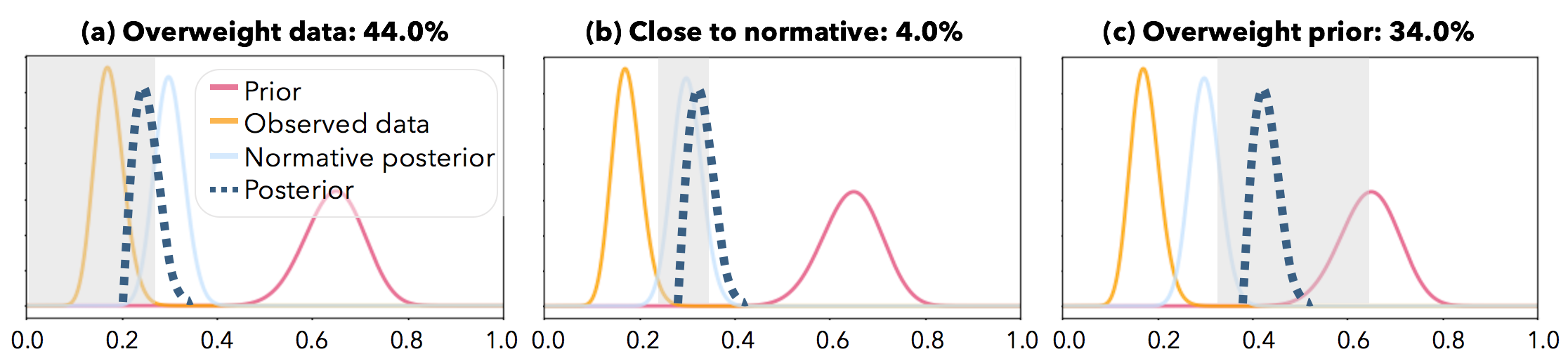}
  \Description{Proportions of participants whose posterior distributions (dotted line) imply overweighting of the mode of the observed data, reasonable alignment with the normative posterior, and overweighting of the mode of the prior distributions.}
 \caption{Proportions of participants whose posterior distributions (dotted line) imply overweighting of the mode of the observed data, reasonable alignment with the normative posterior, and overweighting of the mode of the prior distributions. An additional 18\% of participants (not shown) provided posterior beliefs that were further than the prior from the observed data.}
\vspace{-0.1in}
\label{fig:s1_overshoot}
\end{figure}

Following this observation, we analyzed where each participant's posterior distribution was located relative to the normative posterior distribution (Fig.~\ref{fig:s1_overshoot}). We found that 44\% of participants (22 out of 50) overweighted the mode of the observed data (i.e., their posterior distributions are closer to the observed data than they should be), while 34\% of participants (17 out of 50) overweighted the mode of their prior distribution, and 18\% of the participants (9 out of 50) provided posterior beliefs that moved further than the prior from the observed data. Only two participants (4\%) were within $\pm1\%$ range of the mode of their normative posteriors. 

Per our pre-registration we also report log KL divergence (KLD)~\cite{kullback1951} between normative and observed posteriors. KLD is an information theoretic measure of the difference between two probability distributions.  Examining log KLD at the individual and aggregate levels aligned with our observation from the residual plots: few individuals act ``fully Bayesian'', but in aggregate the responses are close to normative predictions.
The mean log KLD for a participant at the individual level was 0.52 (SD=1.18; 3.31 in non-log terms). 
Normative behavior is represented by a smaller log KLD and non-log KLD close to 0. The aggregate log KLD was -2.18 (non-log KLD=0.11), which aligns with previous work that demonstrates people's collective reasoning is more consistent with Bayesian optimal behaviors even when individuals do not necessarily act as a fully Bayesian agent~\cite{griffiths2006}.

%% file: 4_study2.tex
\section{S2: Elicitation Techniques \& Dataset}
Study 1 (S1) used an elicitation technique from the literature which was designed for fitting Beta distributions to participants' responses using a numerical solution~\cite{fox1966}. While the technique has been shown to be less sensitive to noise from the elicitation process than other techniques~\cite{wu2008}, it is possible that the evidence for approximate or ``sample-based'' Bayesian inference that we observed was an artifact of the elicitation technique. For instance, by asking for a mode value, it is possible that the technique prompted people to consider only a single sample. We are interested in evaluating how robust our result in Study 1 is to changes in the dataset that is presented. 
In a second pre-registered study,\footnote{\url{http://aspredicted.org/blind.php?x=4bf9ci}} we therefore evaluate four different elicitation techniques (including interval elicitation from Study 1) and introduce a new dataset. 
The elicitation techniques vary in the degree to which they ask a participant to provide a full distribution versus a small set of samples. 
By manipulating both representation of uncertainty and the dataset, we aim to gain a better sense of how robust our observation of approximate Bayesian inference is.

\subsection{Developing Elicitation Techniques \& Conditions}
We are interested in comparing a set of interfaces which vary in the format they use to elicit participants' responses. We describe two sample-based techniques of our own design, as well as two elicitation techniques from the literature.
While our data interpretation task requires eliciting a Beta distribution specifically, we expect that the techniques we evaluate should generalize to other symmetric distributions.

\subsubsection{Sample-based Elicitation}
Evidence from research on reasoning with uncertainty (e.g., on classical Bayesian reasoning tasks~\cite{gigerenzer1995}) and uncertainty visualization~\cite{fernandes2018uncertainty,hullman2015,hullman2018,kale2018,kay2016ish} indicates that people are often better at thinking about uncertainty when it is framed as frequency rather than probability.
One way to elicit uncertainty is through a technique that asks people to provide one sample at a time until they have exhausted their mental representation. 
Imagine a person who is asked to provide their expectations for the proportion of women in tech who experience mental issues often. Several possible proportions seem salient to them, including 20\% and 33\%. We devise a sample-based elicitation method that asks a person to articulate a small set of samples (e.g., 5), one at a time (Fig.~\ref{fig:s2_technique}(a)). 

Even if people find it easy to reason in the form of samples, we might still expect that they perceive some samples as more likely. A sample-based elicitation technique would not prevent a person from providing the same sample multiple times, proportional to its expected probability (i.e., resampling with replacement). However, articulating the same sample multiple times can be tedious. For each sample a person provides, our technique asks for a corresponding judgment about the salience of the sample in the form of subjective confidence. Using this technique, the hypothetical person with two samples of 20\% and 33\% might provide 20\% as a first estimate with a higher confidence (e.g., 70 on a scale of 0 to 100), and 33\% as a second estimate with a lower confidence (e.g., 30). In practice, the confidence values do not need to sum to 100 as they can be normalized prior to using them to fit the responses to a distribution.

We created two versions of our sample-based elicitation technique. A \textbf{graphical sample-based elicitation interface} (Fig.~\ref{fig:s2_technique} (a) left) allows participants to provide a predicted value (i.e., sample) by clicking icons in an icon array. This interface is nearly identical to the visual format used to present the observed data, except for the elicitation icon arrays, which use circles. An analogous \textbf{text sample-based elicitation interface} (Fig.~\ref{fig:s2_technique} (a) right) allows participants to provide a predicted value by entering number in a text box. 

As a participant provides their samples, each prior sample is appended to the bottom of the interface so that participants can review their samples and corresponding confidence values before submitting the response.

\subsubsection{Graphical Distribution Elicitation} 
To conduct a Bayesian analysis in many domains (e.g., clinical trials, meteorology, etc.), analysts probe domain experts for uncertainty estimates, then use these to construct a prior distribution~\cite{o2006}. This approach generally assumes that people with domain knowledge possess a relatively complete mental representation of the uncertainty in a parameter. Graphical elicitation of a full subjective distribution has been proposed for use among lay people, such as to elicit preferred retirement outcomes~\cite{goldstein2008}. Recent research indicates that a graphical interface that enables constructing 
a distribution via placing 100 hypothetical outcomes  (``balls'', or circles representing hypothetical outcomes) in multiple ranges (``bins'') allows people to articulate a distribution that they have been presented with more accurately than a method that asks for quantiles of the distribution~\cite{goldstein2014}.
We implemented a \textbf{graphical ``balls and bins'' elicitation interface} (Fig.~\ref{fig:s2_technique}(c)).
Participants are prompted to add exactly 100 balls in bins that span between 0\% to 100\% in increments of 5\% to express the distribution they have in mind.
Relative to the text and graphical sample-based techniques we developed, the graphical balls and bins interface encourages a person to consider their entire subjective probability distribution at once.

\subsubsection{Sample + Partial Distribution Elicitation}
The interval technique we used in Study 1 may be best considered a hybrid approach between approaches that emphasize small sets of samples and those that emphasize a full distribution (Fig.~\ref{fig:s2_technique}(b)). The mode that a participant provides can be thought of as the most salient sample in their priors. The subjective probability that a participant provides is analogous to the probability mass of a partial distribution. 

As in Study 1, participants are first prompted to provide a prediction ($m$). Participants are then asked to provide the subjective probability ($sp$) that the true proportion falls into the range calculated based on the mode value that they entered  ([$m - m*0.25, m + m*0.25$]).\footnote{We elicited two additional random ranges to see how the response is impacted by the ranges. The analysis is in the supplemental material.}

\vspace{-0.1in}
\begin{figure}[htb]
 \centering
  \includegraphics[width=3.5in]{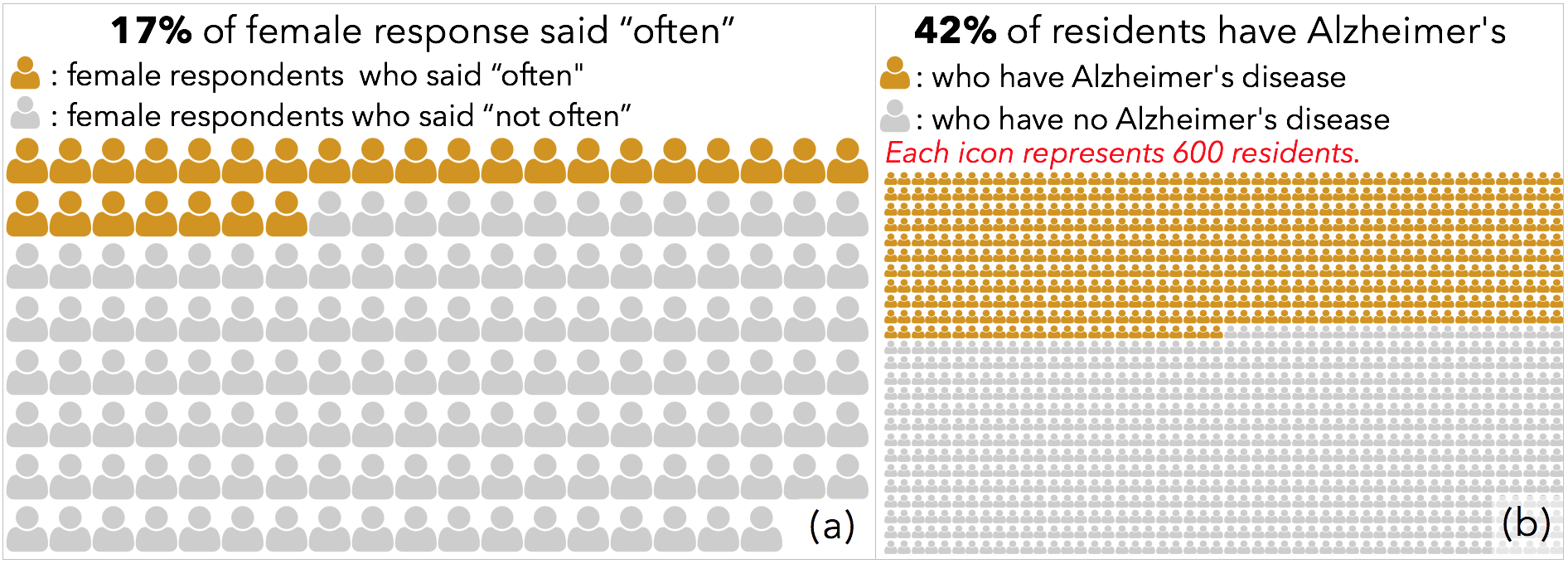}
  \vspace{-0.3in}
  \Description{The data presentations for S1 (a) and S2 (a, b).}
 \caption{The data presentations for S1 (a) and S2 (a, b).}
 \vspace{-0.2in}
\label{fig:vis}
\end{figure}

\subsection{Study Design}

\subsubsection{Dataset and Presentation:}
\begin{figure}[htb]
 \centering
  \includegraphics[width=3.5in]{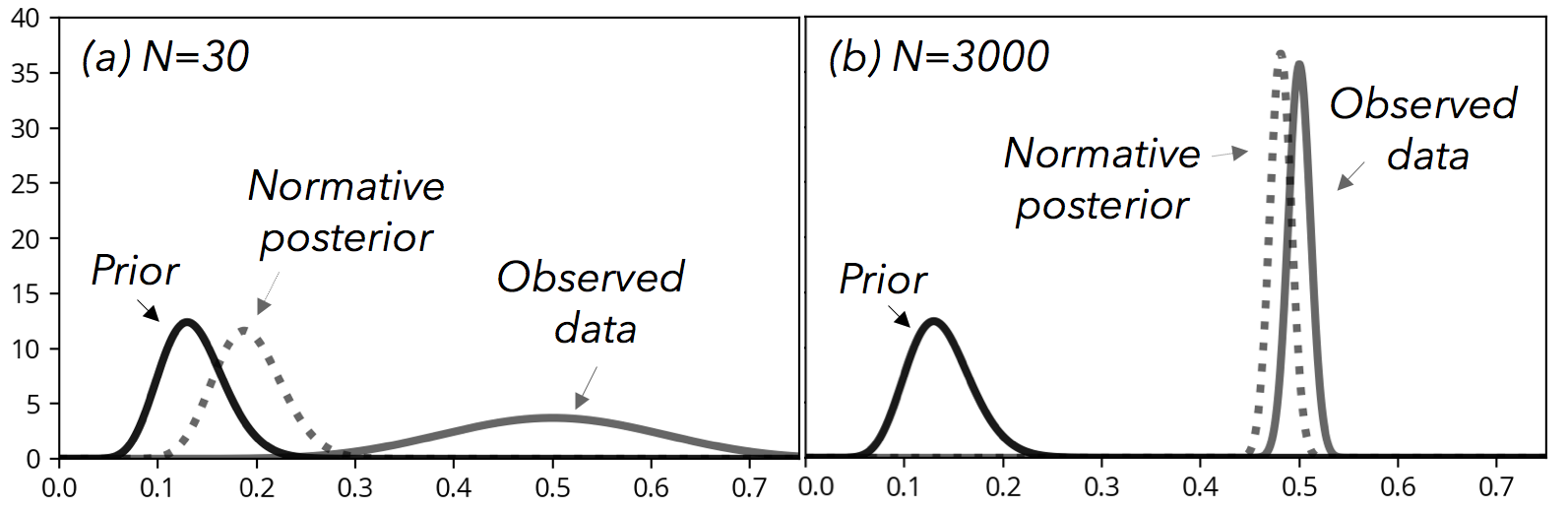}
  \vspace{-0.25in}
  \Description{Illustration of the effect of sample size on normative posteriors given the same prior and observed mode.}
 \caption{Illustration of the effect of sample size on normative posteriors given the same prior and observed mode.}

\label{fig:s2_samplesize}
\end{figure}

We reuse the same proportion dataset used in Study 1 (mental health outcomes among women in the tech industry) and the same icon array visualization. However, we are also interested in understanding how robust our findings are to changes in the nature of the observed data. Specifically, the sample size of the observed data directly influences how closely the normative posterior is expected to align with the data. Intuitively, as the sample size of the observed data increases, the impact of the prior distribution on the normative posterior is reduced. With a very large sample, the normative posterior will be virtually indistinguishable from the data even with a reasonably concentrated prior distribution (Fig.~\ref{fig:s2_samplesize}).

We therefore chose one additional large sample dataset that has been visualized in the New York Times using icon-style visualizations~\cite{nytElderly}. 
This dataset depicts the results of a study of chronic health conditions among assisted living center residents in the U.S. (N=750,000). We chose one type of chronic health condition (Alzheimer's disease or another form of dementia) to formulate our target proportion. We asked participants to reason about ``the proportion of residents who have Alzheimer's disease or another form of dementia'' in the task. We created a visualization (Fig.~\ref{fig:vis} (b)) that shows this proportion in a similar icon array format to that used for the mental health in tech dataset. Because of the size of the sample, we tell participants that each icon represents 600 residents of assisted living centers. 

\begin{figure}[htb]
 \centering
  \includegraphics[width=3.5in]{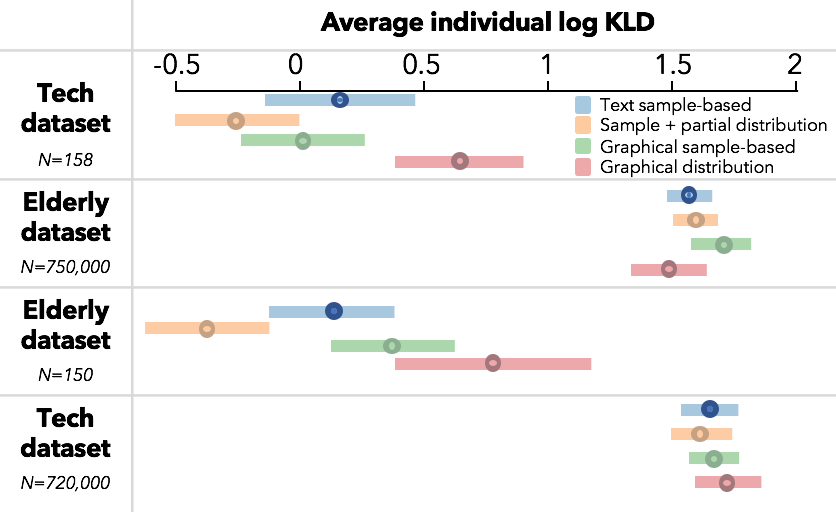}
  \vspace{-0.25in}
  \Description{Bootstrapped 95\% confidence intervals for average log KLDs.}
 \caption{Bootstrapped 95\% confidence intervals for average log KLDs.}
\vspace{-0.1in}
\label{fig:s2_indi_ci}
\end{figure}

\subsubsection{Procedure}
We used the same procedure as in Study 1 (eliciting priors, presenting observed data, eliciting posteriors). However, in Study 2 we randomly assigned participants to one of the four elicitation conditions, and one of the two datasets. On the last page of the experiment, we asked a pre-registered attention-check question about the numeric range in which the observed proportion fell to exclude participants who may not have paid attention to the observed data. Participants were asked to choose an answer among three ranges (0\%-30\%, 30\%-60\%, 60\%-100\%).

\subsubsection{Participants}
Based on a prospective power analysis conducted on pilot data with a desired power of at least 0.8 assuming $\alpha$=0.05, we recruited 800 workers with an approval rating of 98\% or more (400 per dataset, 200 per elicitation condition) in the U.S from Amazon Mechanical Turk (AMT). We disallowed workers who took part in Study 1. We excluded participants who did not respond correctly to our attention check question from the result according to our pre-registration. We posted the task to AMT until 800 participants who correctly answered the attention check question were recruited. Participants received \$1.0 as a reward.  The average time to complete the task was 4.8 minutes (SD=3.35).

\subsection{Results}
\subsubsection{Data Preliminaries}
For each technique, we aimed to use the simple and most direct technique to fit a Beta distribution, so as to minimize noise contributed by the fitting process. 
For sample-based elicitation conditions, we used the Method of Moments~\cite{hansen1982} to estimate distribution parameters (i.e., alpha and beta) using samples provided by each participant. This method provides an estimate using the mean of the samples that participants provided ($\bar{x}$) and the variance of the samples ($\bar{v}$) to calculate beta parameters: $\alpha = \bar{x}(\frac{\bar{x}(1-\bar{x})}{\bar{v}} -1), \beta = (1-\bar{x})(\frac{\bar{x}(1-\bar{x})}{\bar{v}} -1)$. Since we asked participants to provide their subjective confidence with each sample, we calculated weighted $\bar{x}$ by multiplying the value of each sample by the corresponding confidence value. This approach does not provide a unique solution when the participant provides the same values five times or 0 confidence for all samples. In this case, we gave the participant an uninformative uniform prior ($\alpha=1, \beta=1$). We provide a sensitivity analysis to different ways of interpreting these ``deviant''cases in the supplemental material \footnote{https://github.com/yeaseulkim/Bayesian-Visualization}. 
For the graphical distribution condition, we also used the Method of Moments approach by considering each ball as a sample known within a 5\% (the bin width). 
For the sample \& partial distribution condition, we used the same optimization approach we used in Study 1.

\subsubsection{Residual Analysis and Log KLD}
\begin{figure}[t]
 \centering
  \includegraphics[width=3.5in]{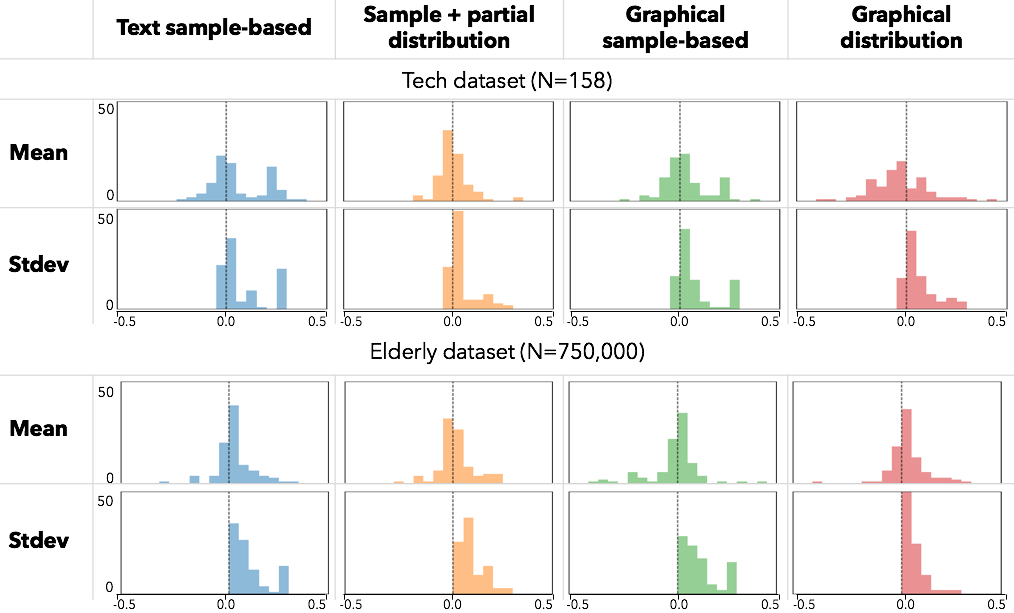}
  \Description{Distributions of residuals (observed-predicted) for participants' posteriors' means and standard deviations and the means and standard deviations of normative posteriors.}
 \caption{Distributions of residuals (observed-predicted) for participants' posteriors' means and standard deviations and the means and standard deviations of normative posteriors.}
\vspace{-0.2in}
\label{fig:s2_diff_mean_stdev}
\end{figure}

To assess the effect of elicitation technique on individual-level alignment with the normative Bayesian solution, we again plot residuals between normative (predicted) means and standard deviations for each participant and observed means and standard deviations (Fig.~\ref{fig:s2_diff_mean_stdev}).
For the tech dataset (N=158) used in Study 2, we observed a similar pattern as in Study 1, with errors roughly equally distributed about zero for means, and around zero but with a slight bias toward the degree of variance of priors (i.e., overestimating variance in the data). 

For the elderly dataset (N=750,000), residuals for means are again roughly symmetric about zero, but residuals for standard deviations are nearly entirely to the right of zero. This suggests a strong tendency for people to be more uncertainty about the true proportion than they should rationally be, given the size of the observed dataset. 

We see some small differences in residual distributions between techniques. For example, those using the graphical balls and bins interface (Fig.~\ref{fig:s2_diff_mean_stdev} fourth column) appear to be slightly more consistent (i.e., more concentrated distribution) and slightly less likely to be biased in their estimates of standard deviation of the elderly dataset (Fig.~\ref{fig:s2_diff_mean_stdev} bottom row). We counted the participants whose responses spanned more than three bins, with the number of balls on either side of the center bin differing by less than two balls.
110 out of 200 participants in this condition attempted to create a symmetric distribution across more than three bins (totaling a 15\% range) for their posterior distribution. Prior work on graphical elicitation has proposed that the axes ranges of an elicitation interface may implicitly influence the predictions that people ``draw''. In the case of the graphical distribution interface, it is possible that participants relied on a heuristic suggesting that distributions should be roughly centered and span more than one bin. 
The small differences in techniques, however, are far less pronounced that the more obvious differences between people's residuals for standard deviation for the (large) elderly dataset versus the (small) tech dataset.  

Per our pre-registration we constructed bootstrapped 95\% confidence intervals for the mean individual log KLD between participants' posteriors and the normative posteriors. Full results from this analysis are in supplemental material. We found that on average, the mean log KLDs from all conditions were larger than we would expect if people are ``fully Bayesian'' at an individual level, further aligning with what we see in Fig.~\ref{fig:s2_diff_mean_stdev}. Across both datasets, we saw no consistent effects of the elicitation techniques on alignment between people's posteriors and the normative posteriors as measured by log KLD. 

To disambiguate whether the difference between the tech dataset and the elderly dataset is due to the different domains of the data or the different sample sizes, we introduced additional datasets by manipulating sample size. We reran the study with the sample sizes switched for the two datasets (tech dataset N=720,000, elderly dataset N=150). We observed the same pattern of results in residual plots (presented in supplemental material), where elicitation techniques did not appear to reliably impact individual's 
residuals in means or standard deviations, but the larger sample size datasets led to residual standard deviations that were strongly biased toward greatly overestimating the amount of uncertainty one should feel given their prior and the observed data.  
We again confirmed these results by examining log KLD (see supplemental material). We speculate that the deviation is caused by a well-documented tendency among people to show insensitivity to sample size and its relationship to variance (sampling error)~\cite{tversky1974}. Hence participants did not weight the value of information captured by the observed elderly dataset as much as they should, given its large sample size (N=750,000).

\begin{figure}[htb]
 \centering
  \includegraphics[width=3.5in]{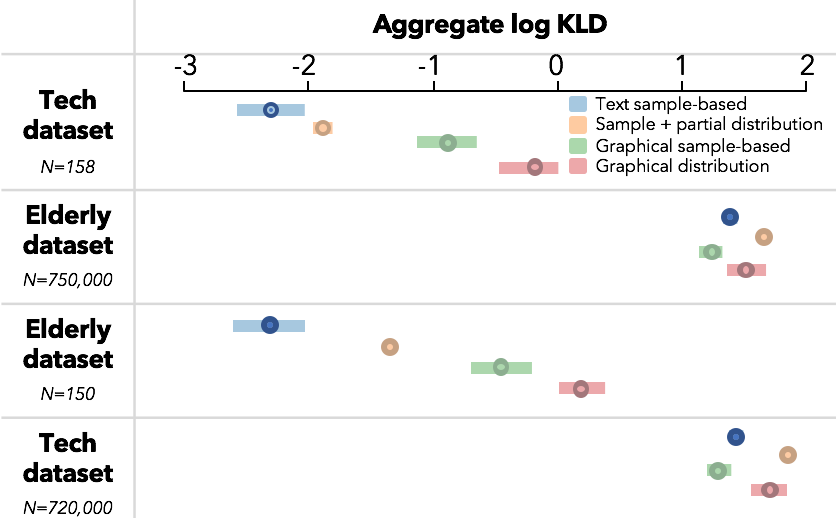}
  \Description{Bootstrapped 95\% confidence intervals for aggregate KLDs.}
 \caption{Bootstrapped 95\% confidence intervals for aggregate KLDs.}
\vspace{-0.1in}
\label{fig:s2_agg_ci}
\end{figure}

We examined the aggregate level log KLD results to 
confirm what the residual plots suggested regarding approximate Bayesian inference for the smaller sample datasets but not for the larger sample datasets. We found that while people's responses were consistent with an approximate or sample-based Bayesian hypothesis for the small sample size datasets, we don't see analogous evidence that people act as sample-based Bayesians for the large sample datasets (Fig.~\ref{fig:s2_agg_ci}(a, b)).

\subsubsection{Perceived Sample Size}
\begin{figure}[htb]
  \begin{center}
    \includegraphics[width=3.5in]{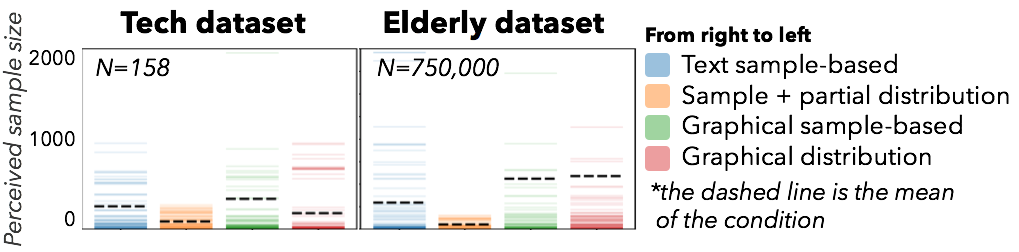}
   \end{center}
   \vspace{-5pt}
   \Description{Perceived sample size as implied by participants' prior and posterior distributions. Participants perceived similar sample sizes between two very different sized datasets.}
      \caption{Perceived sample size as implied by participants' prior and posterior distributions. Participants perceived similar sample sizes between two very different sized datasets.}
    \label{fig:s2_perceivedsample}
    \vspace{-2mm}
\end{figure}

To contextualize these results further, we analyzed perceived sample size.
One benefit of obtaining distributions rather than just expected values (e.g.,~\cite{kim2017, kim2018}) is that we can interpret the parameters of the fitted Beta distributions to gain insight into how participants perceived the data. For a Beta distribution, the two parameters $\alpha$ and $\beta$ are associated  with the sample size that the distribution represents. $\alpha$ stands for the number of successful trials, and $\beta$ stands for the number of failed trials. By treating the participants' posteriors as normative posteriors and using the elicited priors, we reverse-calculated the perceived observed data distribution ($\alpha_\text{perceived data}$ and $\beta_\text{perceived data}$) for each participant (in other words, the counts that a Bayesian would have perceived in the data to arrive at that posterior), then summed these two parameters for sample size. 
Figure ~\ref{fig:s2_perceivedsample} shows how the perceived sample size of the observed data was roughly the same across elicitation techniques and datasets. The mean across all techniques for the tech dataset (N=158) was 212.47 (median=41.14) whereas the mean of elderly dataset (N=750,000) was 359.58 (median=51.51), despite the enormous actual difference in the sample sizes of the observed data (158 vs. 750,000).

%% file: 5_study3.tex
\section{S3: Uncertainty Vis \& Prior Elicitation}
In Study 2 (S2), we observed a pervasive insensitivity to the sample size of the observed data. We turn now to how a Bayesian approach can be used to evaluate how well different visualization alternatives encourage normative interpretations. One natural way to attempt to better calibrate people to the potential for sampling error as a function of sample size is to explicitly present uncertainty information. Visualizing uncertainty may help people make better-calibrated judgments on how much they should weigh the observed data when they incorporate it with their prior beliefs to formulate their posterior beliefs. We also consider the possibility that the insensitivity to large sample sizes that we observe is exaggerated by people anchoring to their prior beliefs because they have been made more salient by elicitation. 
Our goals in our third pre-registered\footnote{\url{http://aspredicted.org/blind.php?x=496ri9}} study are to use Bayesian cognition to evaluate the effect of uncertainty visualization and to better understand the extent to which the act of eliciting people's priors might alter how they update their beliefs.

\subsection{Elicitation Technique \& Dataset } 
\begin{figure}[htb]
 \centering
  \includegraphics[width=3.5in]{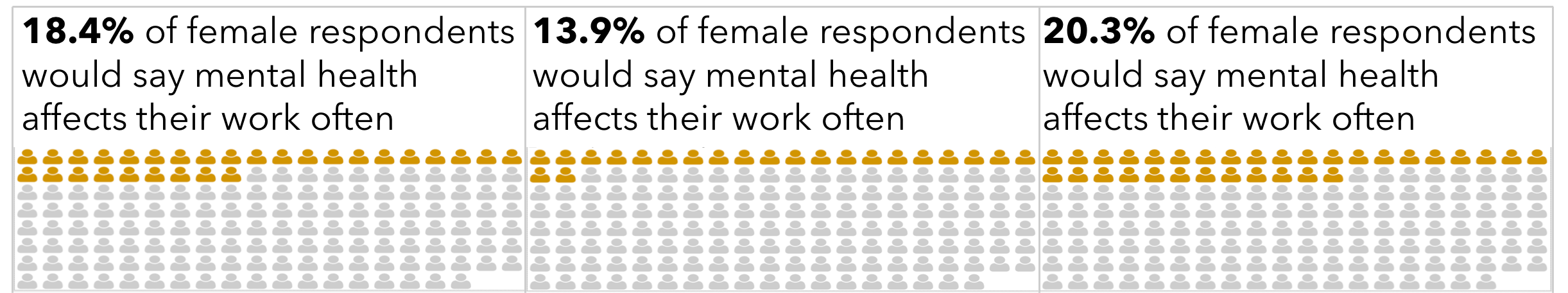}
  \vspace{-0.3in}
  \Description{The example frames from the HOPs (tech dataset).}
 \caption{The example frames from the HOPs (tech dataset).}

\label{fig:s3_hop}
\vspace{-3mm}
\end{figure}

To evaluate our questions, we used the tech dataset (N=158) and the elderly dataset (N=750,000) that we used in Study 2 (Fig.~\ref{fig:vis}). We used the text sample-based technique from Study 2. To show uncertainty around the observed data, we used Hypothetical Outcome Plots (HOPs)~\cite{hullman2015}. HOPs convey uncertainty by animating set outcomes randomly drawn from a target distribution. To create HOPs for each dataset, we constructed a binomial distribution using parameters of the dataset (e.g., $\beta(n=158, p=0.17)$ for the tech dataset), then sampled multiple hypothetical modes from the distribution to present as hypothetical outcomes, using a frame rate of 400ms as suggested by prior work~\cite{hullman2015,kale2018} (Fig.~\ref{fig:s3_hop}).

\vspace{-0.1in}

\subsection{Conditions \& Participants}
\begin{figure}[h]
  
    \includegraphics[width=3.5in]{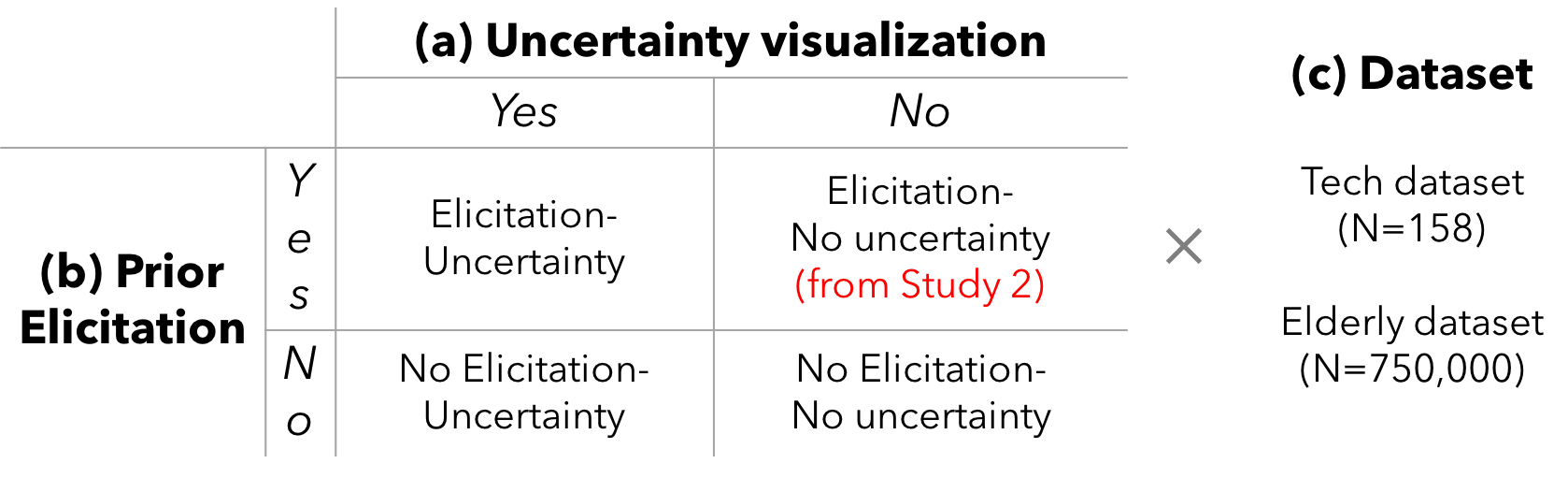}
    \vspace{-0.3in}
   \Description{Table of Study 3 conditions.}
      \caption{Table of Study 3 conditions.}
    \label{fig:s3_condition}
    \vspace{-3mm}
\end{figure}

We crossed two interventions (uncertainty visualization, prior elicitation) to arrive at four study conditions (Fig.~\ref{fig:s3_condition}). Participants in the \textbf{Elicitation-Uncertainty condition} were prompted to externalize their priors before seeing the observed data, then to examine the observed data as HOPs. Participants in the \textbf{Elicitation-No uncertainty condition} were prompted to externalize their priors before seeing the observed data, then to examine the observed data as a static icon array as in Study 2. Participants in the \textbf{No elicitation-Uncertainty condition} were asked to examine the observed data presented with HOPs but were not prompted to externalize their prior beliefs. Lastly, participants in the \textbf{No elicitation-No uncertainty condition} were asked to examine the static observed data without being prompted to externalize their prior beliefs beforehand. Participants in all conditions provided their posterior beliefs after examining the observed data. 

\vspace{-0.15in}

The Elicitation-No uncertainty condition responses consisted of participants' responses from the text sample-based conditions from Study 2 (responses from a total of 200 participants, 100 per dataset). For the remaining conditions, we recruited an additional 600 participants (100 per condition, a total of 300 per dataset) in the U.S from AMT. We disallowed workers who took part in Study 1 or Study 2. We excluded participants who did not respond correctly to our attention check question. We posted the task to AMT until 300 participants who correctly answered the attention check questions were recruited. Participants received \$1.0 as a reward.

\subsection{Analysis Approach}
Per our pre-registration we used a Bayesian linear regression implemented in R's \textit{rethinking} package to evaluate the effect of prior elicitation and uncertainty visualization using a single measure (log KLD). We examined residual plots for mean and variance of participants' posterior distributions for all conditions (see supplemental material) to confirm our model interpretations below. 

To compute the normative posterior for No-elicitation conditions, we used the aggregate priors from participants in the text sample-based condition in Study 2 (Tech dataset: $\alpha=10.79, \beta=18.99$, Elderly dataset:$\alpha=31.25, \beta=39.59$). 
We specified a model to regress the mean effect in individual log KLD on dummy variables indicating whether uncertainty visualization was shown, whether prior elicitation was prompted and which dataset was presented (tech vs. elderly). We specified identical weakly regularizing Gaussian priors for mean effects ($\mu$: 0, $\sigma$: 1) and half-Cauchy priors (Cauchy distributions defined over positive real numbers) for scale parameters ($\mu$: 0, $\sigma$: 1). The thick tailed Cauchy distribution tends to be slightly preferable to Gaussian distributions as a weakly regularizing prior for standard deviations~\cite{mcelreath2018}. 
We included the (mean-centered) time that the participant spent to examine the observed data to interpret whether time spent alone impacted the results.  We present posterior mean estimates of effects with 95\% confidence intervals.

\subsection{Results}
Mean task completion time was 3.8 minutes (SD:2.4) for No-elicitation and 4.7 (SD:3.2) for Elicitation conditions.

\subsubsection{Impacts on Individuals'  Updated Beliefs}
\begin{figure}[htb]
 \centering
  \includegraphics[width=3.5in]{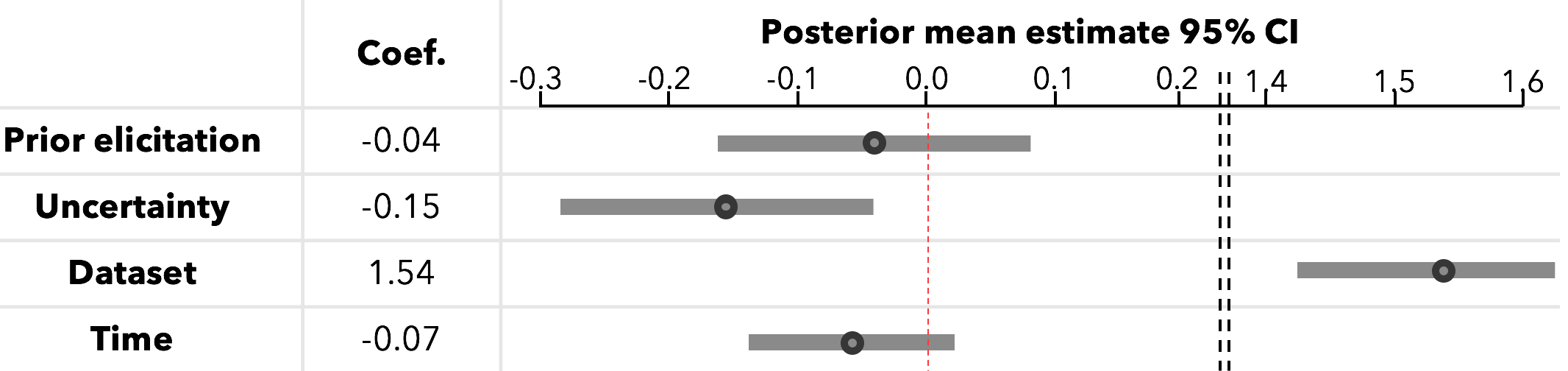}
  \vspace{-0.25in}
  \Description{Posterior mean estimates of effects with 95\% confidence intervals from a model regressing the mean effect on individual log KLD on whether uncertainty visualization was shown, whether prior elicitation was prompted and which dataset was presented.}
 \caption{Posterior mean estimates of effects with 95\% confidence intervals from a model regressing the mean effect on individual log KLD on whether uncertainty visualization was shown, whether prior elicitation was prompted and which dataset was presented. Lower values indicate a greater effect toward lowering log KLD.}

\label{fig:s3_ci}
\end{figure}

Figure~\ref{fig:s3_ci} shows the posterior mean estimates for effects on log KLD. Prior elicitation reduced log KLD of individuals' posterior beliefs relative to the normative Bayesian posteriors but not reliably so (mean:-0.04, 95\% CI:[-0.15,0.1]). Log KLD reliably improved when participants were exposed to uncertainty visualization, with log KLDs relative to the normative posteriors for those who viewed HOPs being on average lower by -0.15 (95\% CI:[-0.29,-0.04]). Being assigned to view the large sample size dataset (i.e., elderly dataset) still had a large impact on results at the individual level, with the average log KLD for those who viewed the large sample dataset being on average 1.54 log KLD units larger than those for the small sample size dataset (95\% CI:[1.42,1.67]). Spending more time examining the observed data reduced log KLD but not reliably (mean=-0.07, 95\% CI:[-0.14,0.01]). 

\vspace{-0.1in}
\vspace{-0.1in}

\subsubsection{Perceived Sample Size}
\begin{figure}[htb]
 \centering
  \includegraphics[width=0.4\textwidth]{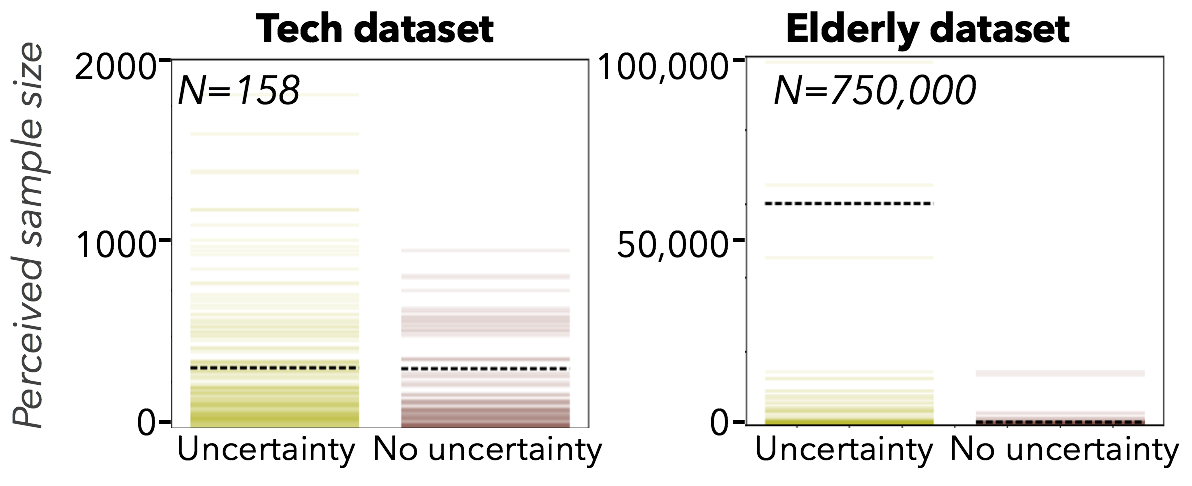}
  \vspace{-0.2in}
  \Description{Perceived sample size for the tech and the elderly datasets.}
 \caption{Perceived sample size for the tech and the elderly datasets. The uncertainty visualization helps participants more accurately perceived sample size in the both datasets.}
\vspace{-0.2in}
\label{fig:s3_perceivedsample}

\end{figure}

Even though participants assigned to examine the large sample size dataset had relatively high log KLDs relative to the small sample size dataset, viewing HOPs did have some impact on how accurately they perceived the sample size of the observed data. Figure ~\ref{fig:s3_perceivedsample} shows how the predicted perceived sample size of the observed data based on the dataset and whether uncertainty (HOPs) was presented. For the tech dataset (N=158), while the means of the No uncertainty and Uncertainty conditions were similar (326.0 vs. 327.3), the median was much closer to the actual sample size of the dataset for the Uncertainty conditions (median perceived: 166.3, true sample size: 158) than the No uncertainty conditions (median perceived: 97.2). 
For the elderly dataset (N=750,000), both the mean and median of the Uncertainty conditions were closer to the true observed sample size (mean perceived: 60,268.9, median perceived: 734.0, true sample size: 750,000) than the No uncertainty conditions (mean perceived: 809.54, median perceived: 216.1).
These results suggest that presenting uncertainty information helps people arrive at normative Bayesian inferences. However, our results also suggest that uncertainty presentation may help some participants more than others.

\vspace{-0.1in}

%% file: 6_discussion.tex
\section{Discussion}
\subsection{Data Interpretation as Bayesian Cognition}
Through three studies, we demonstrated how a Bayesian cognitive modeling approach can be used to interpret and evaluate how people update their beliefs after being exposed to a data presentation. 
Our work represents an important step forward for visualization research for several reasons. 

First, our work provides evidence to suggest that in a naturalistic scenario wherein people bring prior beliefs,  visualization cognition can be interpreted as a Bayesian process. 
While no single study can definitively establish whether people reason about presented data in a Bayesian way, our experiments showed that on average, people's responses were consistent with a sample-based Bayesian account when examining small sample size datasets. On average, people's responses deviated from Bayesian reasoning when presented with large sample size datasets. Future work might test a larger range of sample sizes to further characterize this bias.
We found that the act of eliciting prior beliefs reduces deviation from the normative Bayesian posterior but not reliably so, and that aggregated prior distributions can be used in place of individual level priors to predict a person's deviation from the normative posterior. 


While we used Bayesian cognition as an inferential model to examine how rationally people updated their beliefs at an individual and aggregate level, a Bayesian cognitive model like ours would support studying other influences in visualization or data interpretation. A Bayesian cognitive model could enable designers and researchers to derive more in-depth insights about how graphical, contextual, and individual factors drive the difference between observed versus expected belief change. For example, preferences for certain states of the world are related to, but distinct from, subjective probabilities. Preferences over parameter values (e.g., what should the percentage of women responding positive to the survey question be?) could be elicited as a means of explaining deviation, as could other subjective beliefs like one's trust in the data source.

Our second contribution was to demonstrate how Bayesian cognitive modeling can be used prescriptively to evaluate how well visualization alternatives promote rational behavior. We used the alignment between participants' posteriors and the normative Bayesian posterior to reason about how ``helpful'' it is to add uncertainty representation to a visualization intended for a lay audience. 

An important implication of our findings stems from the evidence we find that visualization users do not necessarily perceive the data they are presented with as ``absolute truth'' with regard to a phenomenon. In the case of very large samples, we found that may deviate substantially from changing their beliefs to match the data. This evidence provides a counterpoint to the implicit assumptions behind many design and evaluation techniques for visualizations.
The prevalence of evaluations that rely on accuracy by comparing interpretations to the data directly and the prevalence of design guidelines that prioritize minimizing perceptual error are more congruent with a normative view in which users \textit{should} match their beliefs to the data than one that emphasizes how visualizations should be used to update users' existing beliefs. 
The Bayesian approach we demonstrate provides a way to quantify how much and in what direction people adjust their beliefs based on new data. Rather than having to carefully craft questions that one thinks might show a difference in accuracy between designs, an evaluator making use of Bayesian models can simply elicit priors and posteriors from a representative sample of users to understand how helpful a visualization is relative to another. 
We demonstrate a graphical sample method that can be used to elicit beliefs for arbitrary visualizations. The statistical literature offers techniques for fitting distributions of varying types to samples (including via Bayesian inference).
By representing beliefs as distributions, as we did in fitting participants' responses to Beta distributions, an evaluator can infer additional information about how users perceived presented data from properties of the fit distributions. For example, we demonstrated how one could calculate the sample size that a rational Bayesian updater would have required to arrive at the same posterior beliefs as the user did, providing a valuable perspective on how a person was off in their interpretation.


\vspace{-0.1in}
\subsection{Eliciting Mental Representation of Uncertainty}
Our work did not identify a single, optimal technique for eliciting an untrained visualization user's internal representation of uncertainty for a proportion. With the exception of the graphical ``balls-and-bins'' interface, which tended to produce posterior beliefs that deviated slightly more from the normative solution, the techniques were difficult to distinguish. However we suspect that 
for more complex datasets and visual representations, the graphical sample-based technique is likely to have advantages.  Imagine observed data visualized as a line chart. A graphical sample-based technique will allow people to simply draw more lines to represent their prior distribution, while other techniques would be more cumbersome (e.g., eliciting intercepts and slopes).  

Our demonstrations involved presenting a proportion statistic. However even this simple scenario required reflecting on the best way to frame the elicitation of a prior. We chose to elicit prior and posterior distributions directly. These distributions are defined over parameter values (i.e., in model space). We chose to change the icons in the icon array format we used for elicitation and presentation to circles, rather than human icons, to better align with the notion of eliciting a population proportion. 
Alternatively, we could have elicited the prior and posterior predictive distributions by asking participants to think about the specific value (e.g., number of women) given some sample size. 

In applying Bayesian modeling to interpretations of more complex datasets and visualizations, we believe that an important consideration will be identifying the appropriate level of ``resolution'' for the prior. For example, if the data is a spatially distributed statistic shown in a choropleth map, the relevant prior might involve subjective distributions over values for individual states, or it may be best represented by distributions over values at a national or regional level. In evaluation scenarios involving Bayesian cognition, the evaluator might consider which prior best matches the intended messages of the data presentation.




The insensitivity toward sample size of the observed data that we reported in Study 2 may partially be because the icon array did not encode sample size directly for the large data. One takeaway is that visualizations should do more to convey sample size, or impacts of sampling error. 

One potential critique of using Bayesian cognition for designing or evaluating visualizations might be that it is unrealistic to expect lay visualization users to possess reasonable (or meaningful) priors on phenomena that tend to be presented to the public in outlets like data journalism. However, we cite multiple forms of evidence to the contrary.
If people did not possess priors or were not able to articulate them, we might expect that with the sample-based elicitation techniques, which required providing samples with confidence, we would see a number of unidentifiable distributions due to all zero confidence values, or no variation in the sample values, for example. 
However, over 85\% of participants who used sample-based techniques provided valid probability distributions. Moreover, across all elicitation techniques we saw that these distributions had non-trivial predictive power for posterior beliefs. A Bayesian model constructed with personal priors achieved a better fit (using Watanabe-Akaike Information Criterion (WAIC) \cite{watanabe2010,gelman2014}) than did a model constructed with the aggregate priors or assuming a uniform prior (WAIC = 2315.7, 3159.8, 3159.8, respectively; see supplemental material for full details).
Finally, several decades worth of work on probability elicitation in large surveys of the public have led economists to believe that laypeople are capable of providing useful information about real world phenomena via subjective probabilities  ~\cite{manski2018}.

\vspace{-0.1in}
\section{Conclusion}
Our work started by asking ``Can Bayesian inference be used to explain how people update their beliefs upon viewing a data presentation?'' In sum, our results demonstrate the promise of using Bayesian cognitive modeling to understand how data presentations like visualizations shape beliefs. Our work demonstrates a path toward better aligning studies of data interpretation with the undeniable effects of prior knowledge, and provides a valuable framework for evaluating new presentation methods.  
\vspace{-0.1in}

\section{Acknowledgements}
We thank Caitlyn McColeman and Nicole Jardine for their
useful feedback.
This work was partially funded by NSF award \#1749266.
This work was also partially supported by the Washington Research Foundation and by a Data Science Environments project award from the Gordon and Betty Moore Foundation (Award \#2013-10-29) and the Alfred P. Sloan Foundation (Award \#3835) to the University of Washington eScience Institute

%% file: main.bbl
\begin{thebibliography}{56}


\ifx \showCODEN    \undefined \def \showCODEN     #1{\unskip}     \fi
\ifx \showDOI      \undefined \def \showDOI       #1{#1}\fi
\ifx \showISBNx    \undefined \def \showISBNx     #1{\unskip}     \fi
\ifx \showISBNxiii \undefined \def \showISBNxiii  #1{\unskip}     \fi
\ifx \showISSN     \undefined \def \showISSN      #1{\unskip}     \fi
\ifx \showLCCN     \undefined \def \showLCCN      #1{\unskip}     \fi
\ifx \shownote     \undefined \def \shownote      #1{#1}          \fi
\ifx \showarticletitle \undefined \def \showarticletitle #1{#1}   \fi
\ifx \showURL      \undefined \def \showURL       {\relax}        \fi
\providecommand\bibfield[2]{#2}
\providecommand\bibinfo[2]{#2}
\providecommand\natexlab[1]{#1}
\providecommand\showeprint[2][]{arXiv:#2}

\bibitem[\protect\citeauthoryear{??}{men}{2016}]%
        {mentalsurvey}
 \bibinfo{year}{2016}\natexlab{}.
\newblock \bibinfo{title}{OSMI Mental Health in Tech Survey 2016}.
\newblock
  \bibinfo{howpublished}{\url{https://www.kaggle.com/osmi/mental-2016\%2011\%2014ealth-in-tech-2016}}.
\newblock
\newblock
\shownote{Accessed: 2018-05-01.}


\bibitem[\protect\citeauthoryear{Alcantara and Esteban}{Alcantara and
  Esteban}{2016}]%
        {gridexample3}
\bibfield{author}{\bibinfo{person}{Chris Alcantara} {and}
  \bibinfo{person}{Chiqui Esteban}.} \bibinfo{year}{2016}\natexlab{}.
\newblock \bibinfo{title}{2016 Election exit polls}.
\newblock \bibinfo{howpublished}{\textit{The Washington Post}, Nov. 29, 2016,
  \url{
  https://www.washingtonpost.com/graphics/politics/2016-election/exit-polls/},}.
\newblock


\bibitem[\protect\citeauthoryear{Benjamin, Rabin, and Raymond}{Benjamin
  et~al\mbox{.}}{2016}]%
        {benjamin2016}
\bibfield{author}{\bibinfo{person}{Daniel~J Benjamin}, \bibinfo{person}{Matthew
  Rabin}, {and} \bibinfo{person}{Collin Raymond}.}
  \bibinfo{year}{2016}\natexlab{}.
\newblock \showarticletitle{A model of nonbelief in the law of large numbers}.
\newblock \bibinfo{journal}{\emph{Journal of the European Economic
  Association}} \bibinfo{volume}{14}, \bibinfo{number}{2}
  (\bibinfo{year}{2016}), \bibinfo{pages}{515--544}.
\newblock


\bibitem[\protect\citeauthoryear{Bloch and Fairfield}{Bloch and
  Fairfield}{2013}]%
        {nytElderly}
\bibfield{author}{\bibinfo{person}{Matthew Bloch} {and} \bibinfo{person}{Hannah
  Fairfield}.} \bibinfo{year}{2013}\natexlab{}.
\newblock \bibinfo{title}{For the Elderly, Diseases That Overlap}.
\newblock \bibinfo{howpublished}{\textit{The New York Times}, Apr 15, 2013,
  \url{https://archive.nytimes.com/www.nytimes.com/interactive/2013/04/16/science/disease-overlap-in-elderly.html},}.
\newblock


\bibitem[\protect\citeauthoryear{Canham and Hegarty}{Canham and
  Hegarty}{2010}]%
        {canham2010}
\bibfield{author}{\bibinfo{person}{Matt Canham} {and} \bibinfo{person}{Mary
  Hegarty}.} \bibinfo{year}{2010}\natexlab{}.
\newblock \showarticletitle{Effects of knowledge and display design on
  comprehension of complex graphics}.
\newblock \bibinfo{journal}{\emph{Learning and instruction}}
  \bibinfo{volume}{20}, \bibinfo{number}{2} (\bibinfo{year}{2010}),
  \bibinfo{pages}{155--166}.
\newblock


\bibitem[\protect\citeauthoryear{Carpenter and Shah}{Carpenter and
  Shah}{1998}]%
        {carpenter1998}
\bibfield{author}{\bibinfo{person}{Patricia~A Carpenter} {and}
  \bibinfo{person}{Priti Shah}.} \bibinfo{year}{1998}\natexlab{}.
\newblock \showarticletitle{A model of the perceptual and conceptual processes
  in graph comprehension.}
\newblock \bibinfo{journal}{\emph{Journal of Experimental Psychology: Applied}}
  \bibinfo{volume}{4}, \bibinfo{number}{2} (\bibinfo{year}{1998}),
  \bibinfo{pages}{75}.
\newblock


\bibitem[\protect\citeauthoryear{Cosmides and Tooby}{Cosmides and
  Tooby}{1996}]%
        {cosmides1996}
\bibfield{author}{\bibinfo{person}{Leda Cosmides} {and} \bibinfo{person}{John
  Tooby}.} \bibinfo{year}{1996}\natexlab{}.
\newblock \showarticletitle{Are humans good intuitive statisticians after all?
  Rethinking some conclusions from the literature on judgment under
  uncertainty}.
\newblock \bibinfo{journal}{\emph{cognition}} \bibinfo{volume}{58},
  \bibinfo{number}{1} (\bibinfo{year}{1996}), \bibinfo{pages}{1--73}.
\newblock


\bibitem[\protect\citeauthoryear{Cox}{Cox}{1999}]%
        {cox1999}
\bibfield{author}{\bibinfo{person}{Richard Cox}.}
  \bibinfo{year}{1999}\natexlab{}.
\newblock \showarticletitle{Representation construction, externalised cognition
  and individual differences}.
\newblock \bibinfo{journal}{\emph{Learning and instruction}}
  \bibinfo{volume}{9}, \bibinfo{number}{4} (\bibinfo{year}{1999}),
  \bibinfo{pages}{343--363}.
\newblock


\bibitem[\protect\citeauthoryear{Fernandes, Walls, Munson, Hullman, and
  Kay}{Fernandes et~al\mbox{.}}{2018}]%
        {fernandes2018uncertainty}
\bibfield{author}{\bibinfo{person}{Michael Fernandes}, \bibinfo{person}{Logan
  Walls}, \bibinfo{person}{Sean Munson}, \bibinfo{person}{Jessica Hullman},
  {and} \bibinfo{person}{Matthew Kay}.} \bibinfo{year}{2018}\natexlab{}.
\newblock \showarticletitle{Uncertainty Displays Using Quantile Dotplots or
  CDFs Improve Transit Decision-Making}. In
  \bibinfo{booktitle}{\emph{Proceedings of the 2018 CHI Conference on Human
  Factors in Computing Systems}}. ACM, \bibinfo{pages}{144}.
\newblock


\bibitem[\protect\citeauthoryear{Fox}{Fox}{1966}]%
        {fox1966}
\bibfield{author}{\bibinfo{person}{Bennett~L Fox}.}
  \bibinfo{year}{1966}\natexlab{}.
\newblock \showarticletitle{A Bayesian approach to reliability assessment}.
\newblock  (\bibinfo{year}{1966}).
\newblock


\bibitem[\protect\citeauthoryear{Gelman, Hwang, and Vehtari}{Gelman
  et~al\mbox{.}}{2014}]%
        {gelman2014}
\bibfield{author}{\bibinfo{person}{Andrew Gelman}, \bibinfo{person}{Jessica
  Hwang}, {and} \bibinfo{person}{Aki Vehtari}.}
  \bibinfo{year}{2014}\natexlab{}.
\newblock \showarticletitle{Understanding predictive information criteria for
  Bayesian models}.
\newblock \bibinfo{journal}{\emph{Statistics and computing}}
  \bibinfo{volume}{24}, \bibinfo{number}{6} (\bibinfo{year}{2014}),
  \bibinfo{pages}{997--1016}.
\newblock


\bibitem[\protect\citeauthoryear{Gigerenzer and Hoffrage}{Gigerenzer and
  Hoffrage}{1995}]%
        {gigerenzer1995}
\bibfield{author}{\bibinfo{person}{Gerd Gigerenzer} {and}
  \bibinfo{person}{Ulrich Hoffrage}.} \bibinfo{year}{1995}\natexlab{}.
\newblock \showarticletitle{How to improve Bayesian reasoning without
  instruction: frequency formats.}
\newblock \bibinfo{journal}{\emph{Psychological review}} \bibinfo{volume}{102},
  \bibinfo{number}{4} (\bibinfo{year}{1995}), \bibinfo{pages}{684}.
\newblock


\bibitem[\protect\citeauthoryear{Goldstein and Gigerenzer}{Goldstein and
  Gigerenzer}{2009}]%
        {goldstein2009}
\bibfield{author}{\bibinfo{person}{Daniel~G Goldstein} {and}
  \bibinfo{person}{Gerd Gigerenzer}.} \bibinfo{year}{2009}\natexlab{}.
\newblock \showarticletitle{Fast and frugal forecasting}.
\newblock \bibinfo{journal}{\emph{International Journal of Forecasting}}
  \bibinfo{volume}{25}, \bibinfo{number}{4} (\bibinfo{year}{2009}),
  \bibinfo{pages}{760--772}.
\newblock


\bibitem[\protect\citeauthoryear{Goldstein, Johnson, and Sharpe}{Goldstein
  et~al\mbox{.}}{2008}]%
        {goldstein2008}
\bibfield{author}{\bibinfo{person}{Daniel~G Goldstein}, \bibinfo{person}{Eric~J
  Johnson}, {and} \bibinfo{person}{William~F Sharpe}.}
  \bibinfo{year}{2008}\natexlab{}.
\newblock \showarticletitle{Choosing outcomes versus choosing products:
  Consumer-focused retirement investment advice}.
\newblock \bibinfo{journal}{\emph{Journal of Consumer Research}}
  \bibinfo{volume}{35}, \bibinfo{number}{3} (\bibinfo{year}{2008}),
  \bibinfo{pages}{440--456}.
\newblock


\bibitem[\protect\citeauthoryear{Goldstein and Rothschild}{Goldstein and
  Rothschild}{2014}]%
        {goldstein2014}
\bibfield{author}{\bibinfo{person}{Daniel~G Goldstein} {and}
  \bibinfo{person}{David Rothschild}.} \bibinfo{year}{2014}\natexlab{}.
\newblock \showarticletitle{Lay understanding of probability distributions.}
\newblock \bibinfo{journal}{\emph{Judgment \& Decision Making}}
  \bibinfo{volume}{9}, \bibinfo{number}{1} (\bibinfo{year}{2014}).
\newblock


\bibitem[\protect\citeauthoryear{Griffiths, Kemp, and Tenenbaum}{Griffiths
  et~al\mbox{.}}{2008}]%
        {griffiths2008}
\bibfield{author}{\bibinfo{person}{Thomas~L Griffiths},
  \bibinfo{person}{Charles Kemp}, {and} \bibinfo{person}{Joshua~B Tenenbaum}.}
  \bibinfo{year}{2008}\natexlab{}.
\newblock \showarticletitle{Bayesian models of cognition}.
\newblock  (\bibinfo{year}{2008}).
\newblock


\bibitem[\protect\citeauthoryear{Griffiths, Lieder, and Goodman}{Griffiths
  et~al\mbox{.}}{2015}]%
        {griffiths2015rational}
\bibfield{author}{\bibinfo{person}{Thomas~L Griffiths}, \bibinfo{person}{Falk
  Lieder}, {and} \bibinfo{person}{Noah~D Goodman}.}
  \bibinfo{year}{2015}\natexlab{}.
\newblock \showarticletitle{Rational use of cognitive resources: Levels of
  analysis between the computational and the algorithmic}.
\newblock \bibinfo{journal}{\emph{Topics in Cognitive Science}}
  \bibinfo{volume}{7}, \bibinfo{number}{2} (\bibinfo{year}{2015}),
  \bibinfo{pages}{217--229}.
\newblock


\bibitem[\protect\citeauthoryear{Griffiths and Tenenbaum}{Griffiths and
  Tenenbaum}{2006}]%
        {griffiths2006}
\bibfield{author}{\bibinfo{person}{Thomas~L Griffiths} {and}
  \bibinfo{person}{Joshua~B Tenenbaum}.} \bibinfo{year}{2006}\natexlab{}.
\newblock \showarticletitle{Optimal predictions in everyday cognition}.
\newblock \bibinfo{journal}{\emph{Psychological science}} \bibinfo{volume}{17},
  \bibinfo{number}{9} (\bibinfo{year}{2006}), \bibinfo{pages}{767--773}.
\newblock


\bibitem[\protect\citeauthoryear{Hansen}{Hansen}{1982}]%
        {hansen1982}
\bibfield{author}{\bibinfo{person}{Lars~Peter Hansen}.}
  \bibinfo{year}{1982}\natexlab{}.
\newblock \showarticletitle{Large sample properties of generalized method of
  moments estimators}.
\newblock \bibinfo{journal}{\emph{Econometrica: Journal of the Econometric
  Society}} (\bibinfo{year}{1982}), \bibinfo{pages}{1029--1054}.
\newblock


\bibitem[\protect\citeauthoryear{Hegarty}{Hegarty}{2004}]%
        {hegarty2004}
\bibfield{author}{\bibinfo{person}{Mary Hegarty}.}
  \bibinfo{year}{2004}\natexlab{}.
\newblock \showarticletitle{Diagrams in the mind and in the world: Relations
  between internal and external visualizations}. In
  \bibinfo{booktitle}{\emph{International Conference on Theory and Application
  of Diagrams}}. Springer, \bibinfo{pages}{1--13}.
\newblock


\bibitem[\protect\citeauthoryear{Hegarty and Steinhoff}{Hegarty and
  Steinhoff}{1997}]%
        {hegarty1997}
\bibfield{author}{\bibinfo{person}{Mary Hegarty} {and} \bibinfo{person}{Kathryn
  Steinhoff}.} \bibinfo{year}{1997}\natexlab{}.
\newblock \showarticletitle{Individual differences in use of diagrams as
  external memory in mechanical reasoning}.
\newblock \bibinfo{journal}{\emph{Learning and Individual differences}}
  \bibinfo{volume}{9}, \bibinfo{number}{1} (\bibinfo{year}{1997}),
  \bibinfo{pages}{19--42}.
\newblock


\bibitem[\protect\citeauthoryear{Hullman, Kay, Kim, and Shrestha}{Hullman
  et~al\mbox{.}}{2018}]%
        {hullman2018}
\bibfield{author}{\bibinfo{person}{Jessica Hullman}, \bibinfo{person}{Matthew
  Kay}, \bibinfo{person}{Yea-Seul Kim}, {and} \bibinfo{person}{Samana
  Shrestha}.} \bibinfo{year}{2018}\natexlab{}.
\newblock \showarticletitle{Imagining Replications: Graphical Prediction \&
  Discrete Visualizations Improve Recall \& Estimation of Effect Uncertainty}.
\newblock \bibinfo{journal}{\emph{IEEE transactions on visualization and
  computer graphics}} \bibinfo{volume}{24}, \bibinfo{number}{1}
  (\bibinfo{year}{2018}), \bibinfo{pages}{446--456}.
\newblock


\bibitem[\protect\citeauthoryear{Hullman, Resnick, and Adar}{Hullman
  et~al\mbox{.}}{2015}]%
        {hullman2015}
\bibfield{author}{\bibinfo{person}{Jessica Hullman}, \bibinfo{person}{Paul
  Resnick}, {and} \bibinfo{person}{Eytan Adar}.}
  \bibinfo{year}{2015}\natexlab{}.
\newblock \showarticletitle{Hypothetical outcome plots outperform error bars
  and violin plots for inferences about reliability of variable ordering}.
\newblock \bibinfo{journal}{\emph{PloS one}} \bibinfo{volume}{10},
  \bibinfo{number}{11} (\bibinfo{year}{2015}), \bibinfo{pages}{e0142444}.
\newblock


\bibitem[\protect\citeauthoryear{Kaeser}{Kaeser}{2015}]%
        {gridexample2}
\bibfield{author}{\bibinfo{person}{Christopher Kaeser}.}
  \bibinfo{year}{2015}\natexlab{}.
\newblock \bibinfo{title}{A day in the Life}.
\newblock \bibinfo{howpublished}{\textit{The Wall Street Journal}, June 24,
  2015, \url{
  https://www.wsj.com/articles/were-working-more-hoursand-watching-more-tv-1435187603?cb=logged0.2694279181305319},}.
\newblock


\bibitem[\protect\citeauthoryear{Kahneman and Egan}{Kahneman and Egan}{2011}]%
        {kahneman2011}
\bibfield{author}{\bibinfo{person}{Daniel Kahneman} {and}
  \bibinfo{person}{Patrick Egan}.} \bibinfo{year}{2011}\natexlab{}.
\newblock \bibinfo{booktitle}{\emph{Thinking, fast and slow}}.
  Vol.~\bibinfo{volume}{1}.
\newblock \bibinfo{publisher}{Farrar, Straus and Giroux New York}.
\newblock


\bibitem[\protect\citeauthoryear{Kale, Nguyen, Kay, and Hullman}{Kale
  et~al\mbox{.}}{2018}]%
        {kale2018}
\bibfield{author}{\bibinfo{person}{Alex Kale}, \bibinfo{person}{Francis
  Nguyen}, \bibinfo{person}{Matthew Kay}, {and} \bibinfo{person}{Jessica
  Hullman}.} \bibinfo{year}{2018}\natexlab{}.
\newblock \showarticletitle{Hypothetical Outcome Plots Help Untrained Observers
  Judge Trends in Ambiguous Data}.
\newblock \bibinfo{journal}{\emph{IEEE transactions on visualization and
  computer graphics}} (\bibinfo{year}{2018}).
\newblock


\bibitem[\protect\citeauthoryear{Kay, Kola, Hullman, and Munson}{Kay
  et~al\mbox{.}}{2016}]%
        {kay2016ish}
\bibfield{author}{\bibinfo{person}{Matthew Kay}, \bibinfo{person}{Tara Kola},
  \bibinfo{person}{Jessica~R Hullman}, {and} \bibinfo{person}{Sean~A Munson}.}
  \bibinfo{year}{2016}\natexlab{}.
\newblock \showarticletitle{When (ish) is my bus?: User-centered visualizations
  of uncertainty in everyday, mobile predictive systems}. In
  \bibinfo{booktitle}{\emph{Proceedings of the 2016 CHI Conference on Human
  Factors in Computing Systems}}. ACM, \bibinfo{pages}{5092--5103}.
\newblock


\bibitem[\protect\citeauthoryear{Kersten and Yuille}{Kersten and
  Yuille}{2003}]%
        {kersten2003}
\bibfield{author}{\bibinfo{person}{Daniel Kersten} {and} \bibinfo{person}{Alan
  Yuille}.} \bibinfo{year}{2003}\natexlab{}.
\newblock \showarticletitle{Bayesian models of object perception}.
\newblock \bibinfo{journal}{\emph{Current opinion in neurobiology}}
  \bibinfo{volume}{13}, \bibinfo{number}{2} (\bibinfo{year}{2003}),
  \bibinfo{pages}{150--158}.
\newblock


\bibitem[\protect\citeauthoryear{Kim, Reinecke, and Hullman}{Kim
  et~al\mbox{.}}{2017}]%
        {kim2017}
\bibfield{author}{\bibinfo{person}{Yea-Seul Kim}, \bibinfo{person}{Katharina
  Reinecke}, {and} \bibinfo{person}{Jessica Hullman}.}
  \bibinfo{year}{2017}\natexlab{}.
\newblock \showarticletitle{Explaining the gap: Visualizing one's predictions
  improves recall and comprehension of data}. In
  \bibinfo{booktitle}{\emph{Proceedings of the 2017 CHI Conference on Human
  Factors in Computing Systems}}. ACM, \bibinfo{pages}{1375--1386}.
\newblock


\bibitem[\protect\citeauthoryear{Kim, Reinecke, and Hullman}{Kim
  et~al\mbox{.}}{2018}]%
        {kim2018}
\bibfield{author}{\bibinfo{person}{Yea-Seul Kim}, \bibinfo{person}{Katharina
  Reinecke}, {and} \bibinfo{person}{Jessica Hullman}.}
  \bibinfo{year}{2018}\natexlab{}.
\newblock \showarticletitle{Data Through Others' Eyes: The Impact of
  Visualizing Others' Expectations on Visualization Interpretation}.
\newblock \bibinfo{journal}{\emph{IEEE transactions on visualization and
  computer graphics}} \bibinfo{volume}{24}, \bibinfo{number}{1}
  (\bibinfo{year}{2018}), \bibinfo{pages}{760--769}.
\newblock


\bibitem[\protect\citeauthoryear{Kosslyn}{Kosslyn}{1989}]%
        {kosslyn1989}
\bibfield{author}{\bibinfo{person}{Stephen~M Kosslyn}.}
  \bibinfo{year}{1989}\natexlab{}.
\newblock \showarticletitle{Understanding charts and graphs}.
\newblock \bibinfo{journal}{\emph{Applied cognitive psychology}}
  \bibinfo{volume}{3}, \bibinfo{number}{3} (\bibinfo{year}{1989}),
  \bibinfo{pages}{185--225}.
\newblock


\bibitem[\protect\citeauthoryear{Kullback and Leibler}{Kullback and
  Leibler}{1951}]%
        {kullback1951}
\bibfield{author}{\bibinfo{person}{Solomon Kullback} {and}
  \bibinfo{person}{Richard~A Leibler}.} \bibinfo{year}{1951}\natexlab{}.
\newblock \showarticletitle{On information and sufficiency}.
\newblock \bibinfo{journal}{\emph{The annals of mathematical statistics}}
  \bibinfo{volume}{22}, \bibinfo{number}{1} (\bibinfo{year}{1951}),
  \bibinfo{pages}{79--86}.
\newblock


\bibitem[\protect\citeauthoryear{Lewandowsky, Griffiths, and
  Kalish}{Lewandowsky et~al\mbox{.}}{2009}]%
        {lewandowsky2009}
\bibfield{author}{\bibinfo{person}{Stephan Lewandowsky},
  \bibinfo{person}{Thomas~L Griffiths}, {and} \bibinfo{person}{Michael~L
  Kalish}.} \bibinfo{year}{2009}\natexlab{}.
\newblock \showarticletitle{The wisdom of individuals: Exploring people's
  knowledge about everyday events using iterated learning}.
\newblock \bibinfo{journal}{\emph{Cognitive science}} \bibinfo{volume}{33},
  \bibinfo{number}{6} (\bibinfo{year}{2009}), \bibinfo{pages}{969--998}.
\newblock


\bibitem[\protect\citeauthoryear{Manski}{Manski}{2018}]%
        {manski2018}
\bibfield{author}{\bibinfo{person}{Charles~F Manski}.}
  \bibinfo{year}{2018}\natexlab{}.
\newblock \showarticletitle{Survey measurement of probabilistic macroeconomic
  expectations: progress and promise}.
\newblock \bibinfo{journal}{\emph{NBER Macroeconomics Annual}}
  \bibinfo{volume}{32}, \bibinfo{number}{1} (\bibinfo{year}{2018}),
  \bibinfo{pages}{411--471}.
\newblock


\bibitem[\protect\citeauthoryear{Marsh}{Marsh}{2012}]%
        {gridexample1}
\bibfield{author}{\bibinfo{person}{Bill Marsh}.}
  \bibinfo{year}{2012}\natexlab{}.
\newblock \bibinfo{title}{Are We in the Midst Of a Sixth Mass Extinction?}
\newblock \bibinfo{howpublished}{\textit{The New York Times}, June 1, 2012,
  \url{https://archive.nytimes.com/www.nytimes.com/interactive/2012/06/01/opinion/sunday/are-we-in-the-midst-of-a-sixth-mass-extinction.html},}.
\newblock


\bibitem[\protect\citeauthoryear{Mayer}{Mayer}{2014}]%
        {mayer2014}
\bibfield{author}{\bibinfo{person}{Richard~E Mayer}.}
  \bibinfo{year}{2014}\natexlab{}.
\newblock \showarticletitle{Cognitive theory of multimedia learning}.
\newblock \bibinfo{journal}{\emph{The Cambridge handbook of multimedia
  learning}}  \bibinfo{volume}{43} (\bibinfo{year}{2014}).
\newblock


\bibitem[\protect\citeauthoryear{McElreath}{McElreath}{2018}]%
        {mcelreath2018}
\bibfield{author}{\bibinfo{person}{Richard McElreath}.}
  \bibinfo{year}{2018}\natexlab{}.
\newblock \bibinfo{booktitle}{\emph{Statistical Rethinking: A Bayesian Course
  with Examples in R and Stan}}.
\newblock \bibinfo{publisher}{CRC Press}.
\newblock


\bibitem[\protect\citeauthoryear{Munnich, Ranney, and Song}{Munnich
  et~al\mbox{.}}{2007}]%
        {munnich2007}
\bibfield{author}{\bibinfo{person}{Edward Munnich},
  \bibinfo{person}{Micheal~Andrew Ranney}, {and} \bibinfo{person}{Mirian
  Song}.} \bibinfo{year}{2007}\natexlab{}.
\newblock \showarticletitle{Surprise, surprise: The role of surprising
  numerical feedback in belief change}. In
  \bibinfo{booktitle}{\emph{Proceedings of the Annual Meeting of the Cognitive
  Science Society}}, Vol.~\bibinfo{volume}{29}.
\newblock


\bibitem[\protect\citeauthoryear{Natter and Berry}{Natter and Berry}{2005}]%
        {natter2005}
\bibfield{author}{\bibinfo{person}{Hedwig~M Natter} {and}
  \bibinfo{person}{Dianne~C Berry}.} \bibinfo{year}{2005}\natexlab{}.
\newblock \showarticletitle{Effects of active information processing on the
  understanding of risk information}.
\newblock \bibinfo{journal}{\emph{Applied Cognitive Psychology}}
  \bibinfo{volume}{19}, \bibinfo{number}{1} (\bibinfo{year}{2005}),
  \bibinfo{pages}{123--135}.
\newblock


\bibitem[\protect\citeauthoryear{O'Hagan, Buck, Daneshkhah, Eiser, Garthwaite,
  Jenkinson, Oakley, and Rakow}{O'Hagan et~al\mbox{.}}{2006}]%
        {o2006}
\bibfield{author}{\bibinfo{person}{Anthony O'Hagan}, \bibinfo{person}{Caitlin~E
  Buck}, \bibinfo{person}{Alireza Daneshkhah}, \bibinfo{person}{J~Richard
  Eiser}, \bibinfo{person}{Paul~H Garthwaite}, \bibinfo{person}{David~J
  Jenkinson}, \bibinfo{person}{Jeremy~E Oakley}, {and} \bibinfo{person}{Tim
  Rakow}.} \bibinfo{year}{2006}\natexlab{}.
\newblock \bibinfo{booktitle}{\emph{Uncertain judgements: eliciting experts'
  probabilities}}.
\newblock \bibinfo{publisher}{John Wiley \& Sons}.
\newblock


\bibitem[\protect\citeauthoryear{Padilla, Creem-Regehr, Hegarty, and
  Stefanucci}{Padilla et~al\mbox{.}}{2018}]%
        {padilla2018}
\bibfield{author}{\bibinfo{person}{Lace~M Padilla}, \bibinfo{person}{Sarah~H
  Creem-Regehr}, \bibinfo{person}{Mary Hegarty}, {and}
  \bibinfo{person}{Jeanine~K Stefanucci}.} \bibinfo{year}{2018}\natexlab{}.
\newblock \showarticletitle{Decision making with visualizations: a cognitive
  framework across disciplines}.
\newblock \bibinfo{journal}{\emph{Cognitive research: principles and
  implications}} \bibinfo{volume}{3}, \bibinfo{number}{1}
  (\bibinfo{year}{2018}), \bibinfo{pages}{29}.
\newblock


\bibitem[\protect\citeauthoryear{Pandey, Manivannan, Nov, Satterthwaite, and
  Bertini}{Pandey et~al\mbox{.}}{2014}]%
        {pandey2014}
\bibfield{author}{\bibinfo{person}{Anshul~Vikram Pandey},
  \bibinfo{person}{Anjali Manivannan}, \bibinfo{person}{Oded Nov},
  \bibinfo{person}{Margaret Satterthwaite}, {and} \bibinfo{person}{Enrico
  Bertini}.} \bibinfo{year}{2014}\natexlab{}.
\newblock \showarticletitle{The persuasive power of data visualization}.
\newblock \bibinfo{journal}{\emph{IEEE transactions on visualization and
  computer graphics}} \bibinfo{volume}{20}, \bibinfo{number}{12}
  (\bibinfo{year}{2014}), \bibinfo{pages}{2211--2220}.
\newblock


\bibitem[\protect\citeauthoryear{Pinker}{Pinker}{1990}]%
        {pinker1990}
\bibfield{author}{\bibinfo{person}{Steven Pinker}.}
  \bibinfo{year}{1990}\natexlab{}.
\newblock \showarticletitle{A theory of graph comprehension}.
\newblock \bibinfo{journal}{\emph{Artificial intelligence and the future of
  testing}} (\bibinfo{year}{1990}), \bibinfo{pages}{73--126}.
\newblock


\bibitem[\protect\citeauthoryear{Shah, Mayer, and Hegarty}{Shah
  et~al\mbox{.}}{1999}]%
        {shah1999}
\bibfield{author}{\bibinfo{person}{Priti Shah}, \bibinfo{person}{Richard~E
  Mayer}, {and} \bibinfo{person}{Mary Hegarty}.}
  \bibinfo{year}{1999}\natexlab{}.
\newblock \showarticletitle{Graphs as aids to knowledge construction: Signaling
  techniques for guiding the process of graph comprehension.}
\newblock \bibinfo{journal}{\emph{Journal of educational psychology}}
  \bibinfo{volume}{91}, \bibinfo{number}{4} (\bibinfo{year}{1999}),
  \bibinfo{pages}{690}.
\newblock


\bibitem[\protect\citeauthoryear{Stern, Aprea, and Ebner}{Stern
  et~al\mbox{.}}{2003}]%
        {stern2003}
\bibfield{author}{\bibinfo{person}{Elsbeth Stern}, \bibinfo{person}{Carmela
  Aprea}, {and} \bibinfo{person}{Hermann~G Ebner}.}
  \bibinfo{year}{2003}\natexlab{}.
\newblock \showarticletitle{Improving cross-content transfer in text processing
  by means of active graphical representation}.
\newblock \bibinfo{journal}{\emph{Learning and Instruction}}
  \bibinfo{volume}{13}, \bibinfo{number}{2} (\bibinfo{year}{2003}),
  \bibinfo{pages}{191--203}.
\newblock


\bibitem[\protect\citeauthoryear{Steyvers, Tenenbaum, Wagenmakers, and
  Blum}{Steyvers et~al\mbox{.}}{2003}]%
        {steyvers2003}
\bibfield{author}{\bibinfo{person}{Mark Steyvers}, \bibinfo{person}{Joshua~B
  Tenenbaum}, \bibinfo{person}{Eric-Jan Wagenmakers}, {and}
  \bibinfo{person}{Ben Blum}.} \bibinfo{year}{2003}\natexlab{}.
\newblock \showarticletitle{Inferring causal networks from observations and
  interventions}.
\newblock \bibinfo{journal}{\emph{Cognitive science}} \bibinfo{volume}{27},
  \bibinfo{number}{3} (\bibinfo{year}{2003}), \bibinfo{pages}{453--489}.
\newblock


\bibitem[\protect\citeauthoryear{Tenenbaum, Griffiths, and Kemp}{Tenenbaum
  et~al\mbox{.}}{2006}]%
        {tenenbaum2006}
\bibfield{author}{\bibinfo{person}{Joshua~B Tenenbaum},
  \bibinfo{person}{Thomas~L Griffiths}, {and} \bibinfo{person}{Charles Kemp}.}
  \bibinfo{year}{2006}\natexlab{}.
\newblock \showarticletitle{Theory-based Bayesian models of inductive learning
  and reasoning}.
\newblock \bibinfo{journal}{\emph{Trends in cognitive sciences}}
  \bibinfo{volume}{10}, \bibinfo{number}{7} (\bibinfo{year}{2006}),
  \bibinfo{pages}{309--318}.
\newblock


\bibitem[\protect\citeauthoryear{Thomas L.~Griffiths and Kemp}{Thomas
  L.~Griffiths and Kemp}{2012}]%
        {griffiths2012}
\bibfield{author}{\bibinfo{person}{Joshua B.~Tenenbaum Thomas L.~Griffiths}
  {and} \bibinfo{person}{Charles Kemp}.} \bibinfo{year}{2012}\natexlab{}.
\newblock \showarticletitle{Bayesian Inference}.
\newblock In \bibinfo{booktitle}{\emph{The Oxford handbook of thinking and
  reasoning}}, \bibfield{editor}{\bibinfo{person}{Keith~J. Holyoak} {and}
  \bibinfo{person}{Robert~G. Morrison}} (Eds.). \bibinfo{publisher}{Oxford
  University Press}, \bibinfo{address}{Oxford}.
\newblock


\bibitem[\protect\citeauthoryear{Tversky and Kahneman}{Tversky and
  Kahneman}{1971}]%
        {tversky1971}
\bibfield{author}{\bibinfo{person}{Amos Tversky} {and} \bibinfo{person}{Daniel
  Kahneman}.} \bibinfo{year}{1971}\natexlab{}.
\newblock \showarticletitle{Belief in the law of small numbers.}
\newblock \bibinfo{journal}{\emph{Psychological bulletin}}
  \bibinfo{volume}{76}, \bibinfo{number}{2} (\bibinfo{year}{1971}),
  \bibinfo{pages}{105}.
\newblock


\bibitem[\protect\citeauthoryear{Tversky and Kahneman}{Tversky and
  Kahneman}{1974}]%
        {tversky1974}
\bibfield{author}{\bibinfo{person}{Amos Tversky} {and} \bibinfo{person}{Daniel
  Kahneman}.} \bibinfo{year}{1974}\natexlab{}.
\newblock \showarticletitle{Judgment under uncertainty: Heuristics and biases}.
\newblock \bibinfo{journal}{\emph{science}} \bibinfo{volume}{185},
  \bibinfo{number}{4157} (\bibinfo{year}{1974}), \bibinfo{pages}{1124--1131}.
\newblock


\bibitem[\protect\citeauthoryear{Van~Wijk}{Van~Wijk}{2005}]%
        {van2005}
\bibfield{author}{\bibinfo{person}{Jarke~J Van~Wijk}.}
  \bibinfo{year}{2005}\natexlab{}.
\newblock \showarticletitle{The value of visualization}. In
  \bibinfo{booktitle}{\emph{Visualization, 2005. VIS 05. IEEE}}. IEEE,
  \bibinfo{pages}{79--86}.
\newblock


\bibitem[\protect\citeauthoryear{Vul, Goodman, Griffiths, and Tenenbaum}{Vul
  et~al\mbox{.}}{2014}]%
        {vul2014}
\bibfield{author}{\bibinfo{person}{Edward Vul}, \bibinfo{person}{Noah Goodman},
  \bibinfo{person}{Thomas~L Griffiths}, {and} \bibinfo{person}{Joshua~B
  Tenenbaum}.} \bibinfo{year}{2014}\natexlab{}.
\newblock \showarticletitle{One and done? Optimal decisions from very few
  samples}.
\newblock \bibinfo{journal}{\emph{Cognitive science}} \bibinfo{volume}{38},
  \bibinfo{number}{4} (\bibinfo{year}{2014}), \bibinfo{pages}{599--637}.
\newblock


\bibitem[\protect\citeauthoryear{Watanabe}{Watanabe}{2010}]%
        {watanabe2010}
\bibfield{author}{\bibinfo{person}{Sumio Watanabe}.}
  \bibinfo{year}{2010}\natexlab{}.
\newblock \showarticletitle{Asymptotic equivalence of Bayes cross validation
  and widely applicable information criterion in singular learning theory}.
\newblock \bibinfo{journal}{\emph{Journal of Machine Learning Research}}
  \bibinfo{volume}{11}, \bibinfo{number}{Dec} (\bibinfo{year}{2010}),
  \bibinfo{pages}{3571--3594}.
\newblock


\bibitem[\protect\citeauthoryear{Wu, Shih, and Moore}{Wu et~al\mbox{.}}{2008}]%
        {wu2008}
\bibfield{author}{\bibinfo{person}{Yujun Wu}, \bibinfo{person}{Weichung~J
  Shih}, {and} \bibinfo{person}{Dirk~F Moore}.}
  \bibinfo{year}{2008}\natexlab{}.
\newblock \showarticletitle{Elicitation of a beta prior for Bayesian inference
  in clinical trials}.
\newblock \bibinfo{journal}{\emph{Biometrical Journal}} \bibinfo{volume}{50},
  \bibinfo{number}{2} (\bibinfo{year}{2008}), \bibinfo{pages}{212--223}.
\newblock


\bibitem[\protect\citeauthoryear{Wu, Xu, Chang, and Wu}{Wu
  et~al\mbox{.}}{2017}]%
        {wu2017}
\bibfield{author}{\bibinfo{person}{Yifan Wu}, \bibinfo{person}{Larry Xu},
  \bibinfo{person}{Remco Chang}, {and} \bibinfo{person}{Eugene Wu}.}
  \bibinfo{year}{2017}\natexlab{}.
\newblock \bibinfo{title}{Towards a Bayesian Model of Data Visualization
  Cognition}.
\newblock
\newblock


\bibitem[\protect\citeauthoryear{Zacks and Tversky}{Zacks and Tversky}{1999}]%
        {zacks1999}
\bibfield{author}{\bibinfo{person}{Jeff Zacks} {and} \bibinfo{person}{Barbara
  Tversky}.} \bibinfo{year}{1999}\natexlab{}.
\newblock \showarticletitle{Bars and lines: A study of graphic communication}.
\newblock \bibinfo{journal}{\emph{Memory \& Cognition}} \bibinfo{volume}{27},
  \bibinfo{number}{6} (\bibinfo{year}{1999}), \bibinfo{pages}{1073--1079}.
\newblock


\end{thebibliography}
